\documentclass{ucbthesis}
\usepackage{rotating} 
\usepackage{caption}
\usepackage{subcaption}
\usepackage{{booktabs}}
\usepackage{indentfirst}
\usepackage{graphicx}
\usepackage{setspace}
\usepackage{array}
\usepackage{amsmath}
\usepackage{amssymb}
\usepackage{bbm}
\usepackage{float}
\usepackage{subcaption}
\usepackage{chngcntr}

\usepackage{indentfirst}

\begin{document}


\title{On the Efficacy of Shorting Corporate Bonds as a Tail Risk Hedging Solution}
\author{Travis Cable, Amir Mani, Wei Qi, Georgios Sotiropoulos and Yiyuan Xiong}
\degreesemester{Spring}
\degreeyear{2024}
\degree{Master of Financial Engineering}
\chair{Professor Eben Lazarus}
\othermembers{Emmanuel Vallod}
\numberofmembers{2}
\field{Finance}


\maketitle
\copyrightpage


\begin{abstract}

Bond portfolios such as Pimco's PIMIX constitute billions of dollars of fixed income investments that target high, consistent returns for their investors. Corporate Investment Grade (IG) bonds (those rated BBB- to AAA per Standard \& Poor's) typically make up large portions of these portfolios and are considered relatively safe, liquid, low risk assets. However, during recent market crises including the Global Financial Crisis (GFC) as well as the COVID-19 Pandemic, IG bonds experienced equally large or larger drawdowns than similar duration high-yield (HY) bonds.

United States (US) IG bonds typically trade at modest spreads over US Treasuries, reflecting the credit risk tied to a corporation's default potential. During market crises, IG spreads often widen and liquidity tends to decrease, likely due to increased credit risk (evidenced by higher IG Credit Default Index spreads) and the necessity for asset holders like mutual funds to liquidate assets, including IG credits, to manage margin calls, bolster cash reserves, or meet redemptions. These credit and liquidity premia occur during market drawdowns and tend to move non-linearly with the market. The research herein refers to this non-linearity (during periods of drawdown) as downside convexity, and shows that this market behavior can effectively be captured through a short position established in IG Exchange Traded Funds (ETFs).

The following document details the construction of three signals: Momentum, Liquidity, and Credit, that can be used in combination to signal entries and exits into short IG positions to hedge a typical active bond portfolio (such as PIMIX). A dynamic hedge initiates the short when signals jointly correlate and point to significant future hedged return. The dynamic hedge removes when the short position's predicted hedged return begins to mean revert. This systematic hedge largely avoids IG Credit drawdowns, lowers absolute and downside risk, increases annualised returns and achieves higher Sortino ratios compared to the benchmark funds. The method is best suited to high carry, high active risk funds like PIMIX, though it also generalises to more conservative funds similar to DODIX.


All the results are generated in a robust trading framework that incorporates trading costs from historical bid-ask spreads, implements realistic funding costs (20 - 200 bps) and supports gradual position sizing based on traded volumes. The hedged results work well for portfolios up to \$10bln in assets, making this a simple, cost-effective, and powerful credit hedge that large bond fund portfolio managers could implement quickly.
 

\end{abstract}

\begin{frontmatter}


\tableofcontents
\clearpage
\listoffigures
\clearpage
\listoftables

\end{frontmatter}

\pagestyle{headings}


\chapter{Introduction} \label{Ch. 1 - Intro}

\section{Setting the Stage}
Investment Grade (IG) bonds typically trade at tight spreads to US Treasuries in normal market environments. During sharp market downturns like including the Global Financial Crisis (GFC) of 2008–2009 and the crisis triggered in March 2020 by the Covid-19 pandemic, these spreads widened significantly and were highly volatile, partly due to increased probability of default. This increased credit risk coupled with a lack of liquidity (sellers outnumbering buyers) in the market during crisis can lead to IG bonds trading at substantial discounts.

The concept of tail-risk pertains to the risk associated with abnormal (highly improbable) market conditions. These events, occur infrequently and represent the left-most extremities of a probability distribution. Empirical evidence demonstrates that realized returns are not normally distributed, but rather are kurtotic and negatively skewed with fatter tails.

Holding portfolios during events like the GFC and COVID exposes even fixed income investors to large drawdowns. The notion of buying insurance against these tail events is referred to as hedging in finance. Prevailing tail-hedge strategies insure portfolios via rolling over put options or using a portfolio of volatility focused strategies that spans multiple asset classes.

Fortunately, with the rising popularity of IG Exchanged Traded Funds (ETFs) and the Corporate repurchase (repo) market, short selling of bonds has become increasingly accessible and cost-effective. This alignment provides an opportunity (outlined in the following research) for investors to engage in short selling of cash bonds or credit ETFs in times of turbulent market conditions as indicated by high credit risk, poor market liquidity, and large credit return moves. By shorting these instruments investors can set up an effective hedge on their bond portfolios alternative to that of traditional equity options, volatility futures or volatility swap hedges.

\section{Data Sources and Data Back-testing Period} \label{Data Cleaning}
The following work considers hedges to duration neutral credit returns of bond mutual funds like PIMIX \& DODIX using bond ETFs such as LQD \& HYG and or IG CDXs. As such, all these returns series are sourced in data collection, along with Treasury returns to assure duration neutrality. The trading framework also requires bond prices for spreads and traded volumes. Table \ref{Data Sources} below displays the instruments considered as well as the sources of data and the information acquired from each.

\begin{table}[H]
    \centering
    \caption{Data Sources for the Various Financial Instruments Utilized in the Manuscript.}
    \begin{tabular}{>{\centering\arraybackslash}p{1in}>{\centering\arraybackslash}p{2in}>{\centering\arraybackslash}p{1in}}
    \toprule
    \textbf{Financial Instrument} & \textbf{Data Type} & \textbf{Data Source(s)} \\
    \midrule
    LQD \& HYG & Price, Traded Volume, Duration, Dividend-Yield, Option Implied Volatility & Bloomberg \\
    \bottomrule
    PIMIX \& DODIX & Duration, Gross Adjusted Returns & Bloomberg \\
    \bottomrule
    5-Year IG CDX & Time-series of Constituents & Bloomberg \\
    \bottomrule
    IG Bonds & Duration, Spread, Market Value, Traded Volume & Bloomberg, WRDS, TRACE \\
    \bottomrule
    US Treasury Bonds & Coupon, Duration, Maturity, Yield, Return & Bloomberg, US Dept. of the Treasury \\
    \bottomrule
    \end{tabular}
    \label{Data Sources}
\end{table}

Figure \ref{Hedge Volumes} below displays daily traded volumes for both LQD and HYG from 2010 to 2024. Note that LQD and HYG didn't consistently trade more than a million shares a day on average until late 2018 and 2016, respectively. Consider a situation where a fund manager with \$500mln in assets seeks to hedge their entire portfolio by shorting only LQD at $\approx \$100$ per share and starting at the beginning of 2012. If the manager of the fund was concerned with ``moving the market" and targeted a maximum position taking of 10\% of the average daily traded volume per day, the fund would have required $\approx 35$ trading days to fully accrue the hedged position. In today's market, and due to the increased trading activity in ETFs, the same manager and under the same constraints, would accumulate the full hedged position in just two trading days.

\begin{figure}[H]
  \centering
  \includegraphics[width=1.0\linewidth]{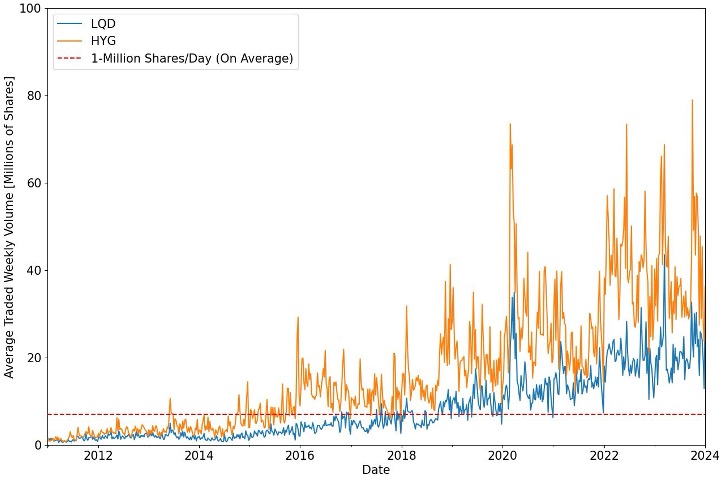}
  \caption{LQD and HYG Weekly Average Traded Volume from 2010 - 2024.}
  \label{Hedge Volumes}
\end{figure}

The current back-testing framework is limited from July 2013 to October 2022, based on availability of required data from Bloomberg, WRDS, and Trace (displayed in Table \ref{Data Sources} above). This is a near full decade of a back-test period. Having established above that ETF liquidity was limited prior to 2013 as visible in Figure \ref{Hedge Volumes}, the back-test history imposed by data limitations aligns well with and reflects times where establishing a moderately large short ETF position starts to become realistic.

\clearpage
\chapter{Literature Review} \label{Ch. 2 - Lit Review}

\section{IG Bond Behavior During Crises}
During crisis in financial market, institutional investors often need to unwind their portfolios to reduce exposures and leverage and could face recurring and uncertain liquidity needs to meet short-term obligations \cite{brown2009scholes}. However, liquidity could dry up in certain markets such as the securitized bond market during the GFC. Most market participants facing a liquidation problem would not sell the illiquid assets but, instead, resort to selling their most liquid assets such as investment grade corporate bonds, as suggested by Scholes \cite{scholes2000crisis}. For example, during the final quarter of 2007, mutual funds decreased their corporate bonds holdings by 15\%, representing 6\% of their total holdings. Simultaneously, they reduced their holdings of illiquid securitized bonds by 9\%, which accounted for 1.9\% of their total holdings, resulting in sharply increasing bond yield spreads and declining prices \cite{manconi2012role}.

\subsection{Global Financial Crisis}
When delinquent subprime mortgages caused the securitized bond market 
to turn illiquid and the resale value of securitized bonds plummeted in August 2007, institutional investors such as bond mutual funds holding securitized bonds (especially those with negative contemporaneous flows, high turnover, or high flow volatility) liquidated more corporate bonds compared to unexposed investors holding same-issuer bonds to satisfy liquidity needs. This greatly contributed to the propagation of crises from the securitized bond market to the corporate bond market, as Mancori et al.\cite{manconi2012role} found. Amid the year long global credit crisis and U.S. recession, IG corporate bond spreads rose to a record 656 basis points in December 2008\cite{reuters2008}.

\subsection{Covid-19 Pandemic}

The corporate bond market experienced a notable expansion since the global financial crisis from approximately \$5.5 trillion in 2008 to an estimated \$9.6 trillion by 2019, where the IG bond sector is about six times the size of the HY bond sector\cite{o2021anatomy}. While insurance companies maintain their position as the primary holders in corporate bond market, mutual funds demonstrated a notable escalation in their holdings, surpassing \$2.2 trillion by the first quarter of 2020. This structural change dramatically increases the demand for liquidity, surpassing the capacity of markets to adequately fulfill this demand during periods of market stress \cite{liang2020corporate}. Liang et al. \cite{liang2020corporate} elucidate the disruptions observed in the corporate bond market where the corporate bonds spreads widened to a greater extent for investment-grade bonds compared to high-yield bonds, which is surprising as high-yield bonds are typically considered riskier and exhibit higher sensitivity to a decline in the economic forecast. Additionally, the rise in bond spreads surpassed the increase in credit-default-swaps (CDS) for the same investment-grade firm, yet this trend was not observed for high-yield bonds \cite{haddad2021selling}. During several instances, the United States Federal Reserve (the Fed) opted to buy vast amounts of corporate bonds to ease the large decline in bond prices. These actions significantly lowered risk spreads and improved market functioning. The key driver behind the dysfunction is the very substantial redemptions in IG corporate bond mutual funds in March 2020 \cite{liang2020corporate}. During the week of March 15th, IG corporate bond ETFs listed in the US were trading at an average discount of 3.36\% compared to their Net Asset Value (NAV) and discounts of over 7\% were seen at times. The divergence could potentially create arbitrage opportunities. 

These works and findings lay the foundation for the work herein, mainly, investigating the efficacy of shorting corporate bonds as a tail risk hedging solution.

\section{IG Bond Factor Models}

An extensive literature works on the cross-sectional risk factors of stock returns. Over the past decades, researchers also have identiﬁed a large number of risk factors that explain the cross-sectional variation in corporate bond returns. Bai et al. \cite{bai2019common} find that downside risk, credit risk, and liquidity risk could assist in predicting future bond returns. The representation for each risk factor may be not straightforward. For example, the liquidity factor lacks a singular variable for direct measurement. As a result, researchers predominantly relied on employing surrogate indicators of market liquidity, notably bid-ask spreads and trading volume. These proxies offer valuable insights into liquidity dynamics, yet they may not comprehensively encapsulate all facets of liquidity, such as market depth or the efficacy of executing substantial trades without inducing price fluctuations. Futhermore, Houwelling et al. \cite{houweling2017factor} provide empirical evidence that portfolios based on factors such as size, low-risk, value, and momentum exhibit economically significant and statistically meaningful alphas within the corporate bond market. 

Li et al.\cite{li2022review} summarize the influencing factors of credit spreads including risk-free interest rate, interest rate risk, liquidity risk, monetary policy, capital structure, corporate governance, bond issuance period, and bond credit rating. Gergana et al. \cite{jostova2013momentum} documents significant momentum in a comprehensive sample of more than 80,000 U.S. corporate bonds. Mueller \cite{mueller2000simple} finds that the growth rate of GDP and the volatility of stock could impact credit spreads on corporate bonds using prices of Delta Airlines' bonds in addition to the leverage ratio of the firm and the riskfree rate of interest. In additiona, many researchers work on bond return predictions based on various factors. Bianchi et al. \cite{bianchi2021bond} show that machine learning methods such as neural networks with non-linear activation functions can lead to more accurate bond return predictability. Kelly et al. \cite{kelly2023modeling} propose a conditional model of corporate bonds returns based on IPCA that utilizes bond and ﬁrm characteristics and show that the trading strategies derived from the model are proﬁtable after net of trading costs. 

The following research focuses on three of the factors (Liquidity, Credit and Momentum) identified by the literature, but in a somewhat novel way.

\clearpage
\chapter{Hedge Signals Design and Implementation} \label{Ch. 3 - Signals}

Chapter \ref{Ch. 2 - Lit Review} details some of the literature on bond return factor models, several of which include factors for probability of default, liquidity and momentum (amongst others). The following section details the construction of these three ``factors" (Credit Risk, Liquidity and Momentum) for the analysis herein.

\section{Credit Risk} \label{Credit Risk Signal}

\subsection{Derivation of Risk-Neutral Distributions}
As mentioned, the current research aims to provide a dynamic hedge for a bond-based portfolio by achieving short exposure to the bond market via ETFs such as LQD, HYG and JNK. Many researchers have incorporated metrics for the probability of default into bond return models, this typically referred to as ``credit" or ``carry" risk \cite{bai2019common}. Instead of attempting to determine individual bonds default probabilities, the approach herein utilizes risk-neutral probability distribution derived from options \cite{malz2014simple} written on LQD. These risk neutral distributions, derived for LQD specifically, reflect the market's forward looking expectation for the IG bond market, and thus should provide some information on future movements.

As in \cite{malz2014simple}, define the value of a call option at time, strike and time until expiration (i.e. tenor), $t, X \text{ and } \tau$, respectively, to be

\begin{equation}\label{CR1}
    c(t,X,\tau) = e^{-r_t\tau} \tilde{\mathbf{E}}_t[\text{max}(S_T - X,0)] = e^{-r_t\tau} \int_X^\infty (s-X)\tilde{\pi}_t(s) ds
\end{equation}

\noindent Equation (\ref{CR1}) represents the risk-neutral expectation of the option price, continuously discounted to the present at the rate $r_t$. Note that $\tilde{\pi}_t(s)$ represents the probability of the underlying asset being $s$ at the time of exercise. In addition, the bounds on the integrand run from $X \text{ to } \infty$ as the call option is out of the money below the strike price $X$ and thus has zero value. Differentiate Eq. (\ref{CR1}) w.r.t. the strike price to obtain

\begin{equation}\label{CR2}
    \frac{\partial}{\partial X}c(t,X,\tau) = e^{-r_t\tau} \frac{\partial}{\partial X}  \bigg\{\int_X^\infty (s-X)\tilde{\pi}_t(s) ds \bigg\} = - e^{-r_t\tau} \int_X^\infty \tilde{\pi}_t(s) ds
\end{equation}

\noindent As $\tilde{\pi}_t(s)$ is a valid probability distribution, it holds that

\begin{equation}\label{CR3}
    \int_{-\infty}^\infty \tilde{\pi}_t(s) ds = \int_0^X \tilde{\pi}_t(s) ds + \int_X^\infty \tilde{\pi}_t(s) ds = 1
\end{equation}

\noindent Rearrange Eq. (\ref{CR3}) and insert into Eq. (\ref{CR2}) to obtain

\begin{equation}\label{CR4}
    \frac{\partial}{\partial X}c(t,X,\tau) = - e^{-r_t\tau} \int_X^\infty \tilde{\pi}_t(s) ds = e^{-r_t\tau} \Bigg[\int_0^X \tilde{\pi}_t(s) ds - 1\Bigg]
\end{equation}

\noindent Multiply both sides by $e^{r_t\tau}$ to see that the risk-neutral cumulative distribution function for the underlying price ($s$) can be written as

\begin{equation}\label{CR5}
    \tilde{\Pi}_t(X) = \int_0^X \tilde{\pi}_t(s) ds = 1 + e^{r_t\tau} \frac{\partial}{\partial X}c(t,X,\tau)
\end{equation}

\noindent Knowing how the cumulative distribution function relates to the probability density function, differentiate Eq. (\ref{CR5}) w.r.t the strike price $X$ to obtain

\begin{equation}\label{CR6}
    \tilde{\pi}_t(X) = \frac{\partial}{\partial X}\tilde{\Pi}_t(X) = e^{r_t\tau} \frac{\partial^2}{\partial X^2}c(t,X,\tau)
\end{equation}

Recall that central difference approximations (with error on the order of $O(\Delta X^2)$) for the first and second derivatives of an arbitrary continuous function $f$ are:

\begin{equation}\label{CR7}
    \begin{gathered}
        \frac{\partial f}{\partial X} \simeq \frac{f(X+\Delta X) - f(X-\Delta X)}{2\Delta X}\\
        \text{and}\\
        \frac{\partial^2 f}{\partial X^2} \simeq \frac{f(X+\Delta X) -2f(X) + f(X-\Delta X)}{\Delta X^2}
    \end{gathered}
\end{equation}

\noindent Discretizations of Eqs. (\ref{CR5} \& \ref{CR6}) based on these central difference approximations then give the risk neutral-probability distribution function and cumulative distribution function as

\begin{equation}\label{CR8}
    \begin{gathered}
        \tilde{\pi}_t(X) \simeq e^{r_t\tau}\bigg[\frac{c(t,X+\Delta X,\tau) - 2c(t,X,\tau) + c(t,X-\Delta X,\tau)}{\Delta X^2}\bigg]\\
        \text{and}\\
        \tilde{\Pi}_t(X) \simeq 1 + e^{r_t\tau}\bigg[\frac{c(t,X+\Delta X,\tau) - c(t,X-\Delta X,\tau)}{2\Delta X}\bigg]
    \end{gathered}
\end{equation}

At the lowest strike price, forward differences (also with error on the order of $O(\Delta X^2)$) give the PDF and CDF as

\begin{equation}\label{CR9}
    \begin{gathered}
        \tilde{\pi}_t(X) \simeq e^{r_t\tau}\bigg[\frac{2c(t,X,\tau) -5c(t,X+\Delta X,\tau) + 4c(t,X+2\Delta X,\tau) - c(t,X+3\Delta X,\tau))}{\Delta X^3}\bigg]\\
        \text{and}\\
        \tilde{\Pi}_t(X) \simeq 1 + e^{r_t\tau}\bigg[\frac{-3c(t,X,\tau) + 4c(t,X+\Delta X,\tau)-c(t,X+2\Delta X,\tau)}{2\Delta X}\bigg]
    \end{gathered}
\end{equation}

\noindent while at the uppermost strike, backward differences yield

\begin{equation}\label{CR10}
    \begin{gathered}
        \tilde{\pi}_t(X) \simeq e^{r_t\tau}\bigg[\frac{2c(t,X,\tau) -5c(t,X-\Delta X,\tau) + 4c(t,X-2\Delta X,\tau) - c(t,X-3\Delta X,\tau))}{\Delta X^3}\bigg]\\
        \text{and}\\
        \tilde{\Pi}_t(X) \simeq 1 + e^{r_t\tau}\bigg[\frac{3c(t,X,\tau) - 4c(t,X-\Delta X,\tau)+c(t,X-2\Delta X,\tau)}{2\Delta X}\bigg]
    \end{gathered}
\end{equation}

\subsubsection{Call Value Function \& Implied Volatility Data}
Eqs. (\ref{CR8}-\ref{CR10}) rely on numerical derivatives of the call-valuation function $c(t, X, \tau)$. The current analysis utilizes the Black-Scholes call pricing formula which requires the current underlying's spot price ($S_t$), option strike price ($X$), time until maturity ($\tau$), annualized dividend yield ($q_t$) and annualized risk-free rate ($r_t$), i.e.

\begin{equation}\label{CR11}
    \begin{aligned}
        c(t, X, \tau) = S_t e^{-q_t \tau} \Phi\bigg[\frac{\text{log}(\frac{S_t}{X}) + (r_t - q_t +\frac{\sigma^2}{2})\tau}{\sigma \sqrt{\tau}}\bigg] - X e^{-r_t \tau} \Phi\bigg[\frac{\text{log}(\frac{S_t}{X}) + (r_t - q_t -\frac{\sigma^2}{2})\tau}{\sigma \sqrt{\tau}}\bigg]
    \end{aligned}
\end{equation}

\noindent and where $\Phi$ is the normal cumulative distribution function.

\begin{figure}[H]
  \centering
  \includegraphics[width=0.75\linewidth]{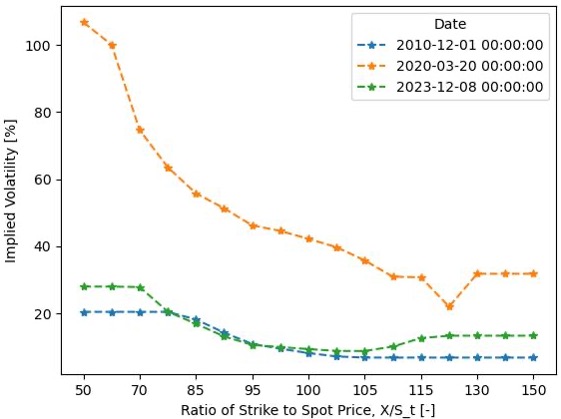}
  \caption{LQD 3-Month Call Option Implied Volatilities versus Strike Price and Collected from Bloomberg.}
  \label{Implied Vols}
\end{figure}

Bloomberg, as mentioned in Section \ref{Data Cleaning}, delivers the data for the daily spot (closing) price, annualized dividend yield and implied volatility for several in-the-money (ITM) and out-of-the-money (OTM) strikes and for several different ETFs. Table \ref{Bloomberg Implied Vol} and Figure \ref{Implied Vols} display implied volatility data for LQD 3-month call options on three different dates.\\

\begin{table}[H]
    \centering
    \caption{Implied Volatilities ($\sigma$) for Call Options of Various Strikes on LQD and Expiring in 3-Months Time and for Three Separate Dates.}
    \begin{tabular}{cccc}
    \toprule
    $\big(\frac{X}{S_t}\big) \cdot 100$ & 2010-12-01 & 2020-03-20 & 2023-12-08 \\
    \midrule
    50 & 0.204990 & 1.067751 & 0.280254 \\
    60 & 0.204990 & 1.000759 & 0.280254 \\
    70 & 0.204990 & 0.746763 & 0.278946 \\
    80 & 0.204990 & 0.635701 & 0.205594 \\
    85 & 0.182460 & 0.558266 & 0.169626 \\
    90 & 0.143122 & 0.513561 & 0.131816 \\
    95 & 0.109516 & 0.462737 & 0.106630 \\
    97.5 & 0.095280 & 0.446174 & 0.100171 \\
    100 & 0.082167 & 0.422495 & 0.094027 \\
    102.5 & 0.072120 & 0.397621 & 0.088550 \\
    105 & 0.068770 & 0.358279 & 0.088269 \\
    110 & 0.068706 & 0.309786 & 0.101580 \\
    115 & 0.068706 & 0.307821 & 0.126621 \\
    120 & 0.068706 & 0.220473 & 0.133911 \\
    130 & 0.068706 & 0.318471 & 0.133911 \\
    140 & 0.068706 & 0.318471 & 0.133911 \\
    150 & 0.068706 & 0.318471 & 0.133911 \\
    \bottomrule
    \end{tabular}
    \label{Bloomberg Implied Vol}
\end{table}

Figure \ref{Implied Vols} shows that for the dates in 2010 and 2023, the implied volatilities are nearly constant for the deep OTM and ITM strikes, while in 2023 and during the uncertainty of the COVID pandemic, the volatility increases exponentially at the lower ITM strikes. This behavior poses an interesting challenge when trying to fit the data. Malz \cite{malz2014simple} assumes that the volatilities after 80 and 120\% are constant and fits a clamped spline (first derivative equal to zero at the boundaries) to the daily volatility versus strike price data. 

Rather than assuming a certain form of the implied volatility function a priori, an algorithm sweeps through the time-series of implied volatilities (day by day), selecting appropriate boundary conditions (BCs) and fitting cubic splines to the daily volatility versus strike function. For example, consider the volatility data listed in Table \ref{Bloomberg Implied Vol} and for the date ``2010-12-01". The algorithm fits a spline with clamped (i.e. 0 first derivative) BCs with the endpoints being 80 and 105\%, since the volatilites are constant further into and out of the money, respectively, after these strikes. In contrast, for the date in 2020, the algorithm would apply the clamped BCs at 50 \& 130\% since the volatilites are still increasing at the bottom end of the curve and are approximately constant after 130\% at the top end.

\begin{figure}[H]
  \centering
  \includegraphics[width=1.0\linewidth]{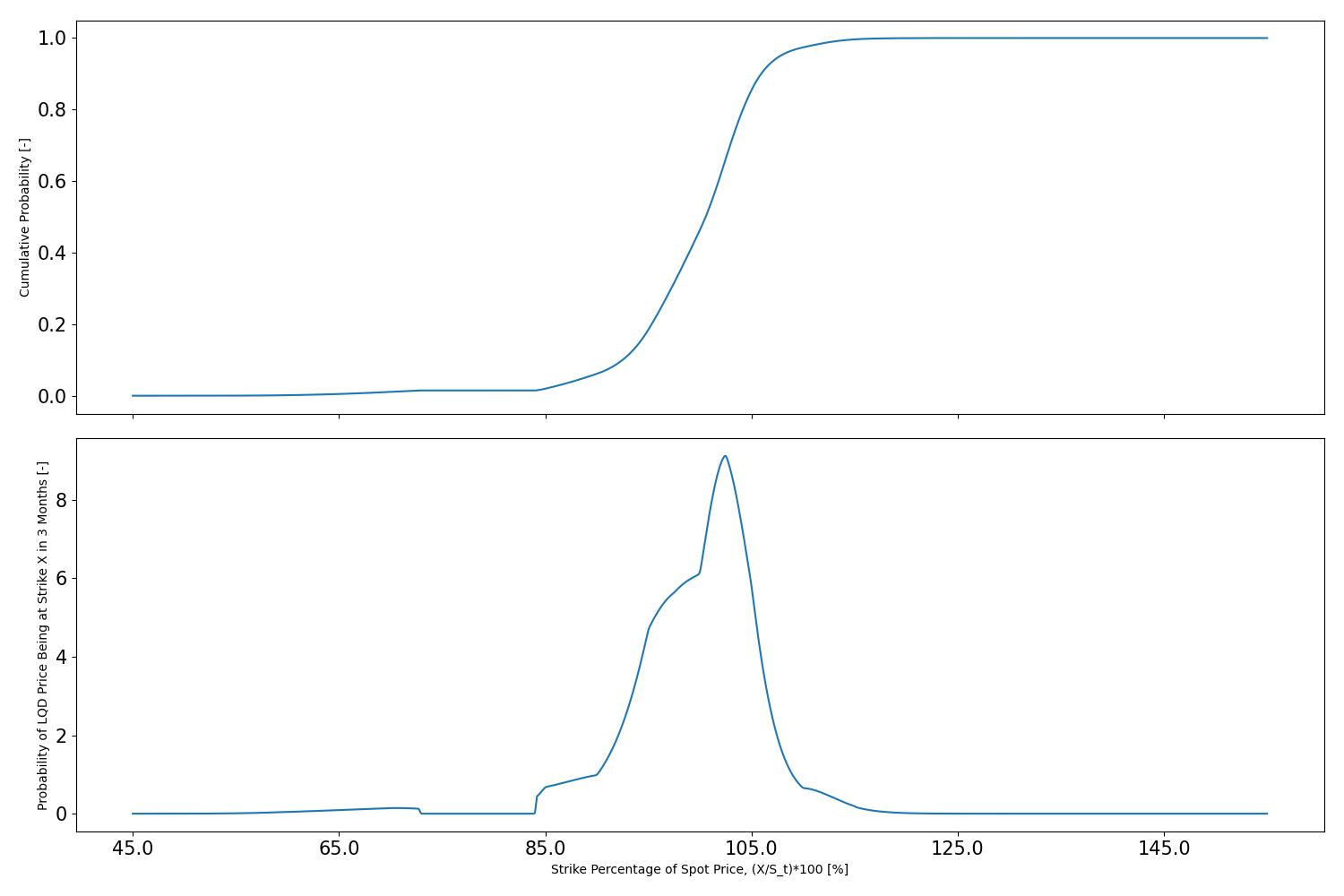}
  \caption{Risk-Neutral CDF and PDF for LQD on 03-23-2023 Derived from Daily Call Option Implied Volatilities.}
  \label{Distributions}
\end{figure}

With the volatility functions (splines) in hand, Eq. \ref{CR11} determines the call value function for each day, from which Eqs. (\ref{CR8}-\ref{CR10}) render the daily CDFs. Despite obtaining more ITM and OTM volatility data (as compared to \cite{malz2014simple}), some CDFs are still noisy and need further cleaning. To navigate this issues, when calculating the CDFs, a numerical procedure ensures the properties of the CDF (i.e. monotonously increasing and bounded between 0 and 1). Figure \ref{Distributions} displays the resulting PDF and CDF for another randomly selected date (March 23rd, 2023). The CDF behaves well while the PDF is slightly distorted in the left tail for this particular date.

\subsubsection{Removing Duration Effects and Final Credit Risk Time-Series}
The daily CDFs, being well defined (being related to the first numerical derivative of the call value function), assist in removing duration effects from the LQD probability of drawdown. The price of LQD, being an ETF composed of IG bonds, is subject to interest rate movements as well as other factors (momentum, spreads, default probability, etc.). To remove the effect of interest rate movements from the predicted drawdowns of LQD, consider first the risk-neutral probability distributions derived from an ETF composed primarily of the US Treasury bonds with maturities between seven to ten years (ticker: IEF). IEF has an effective duration similar to that of LQD, and as such, can be utilized to remove the duration component from the LQD drawdowns since its movements are almost completely driven by interest rate movements.

\begin{figure}[H]
  \centering
  \includegraphics[width=1.0\linewidth]{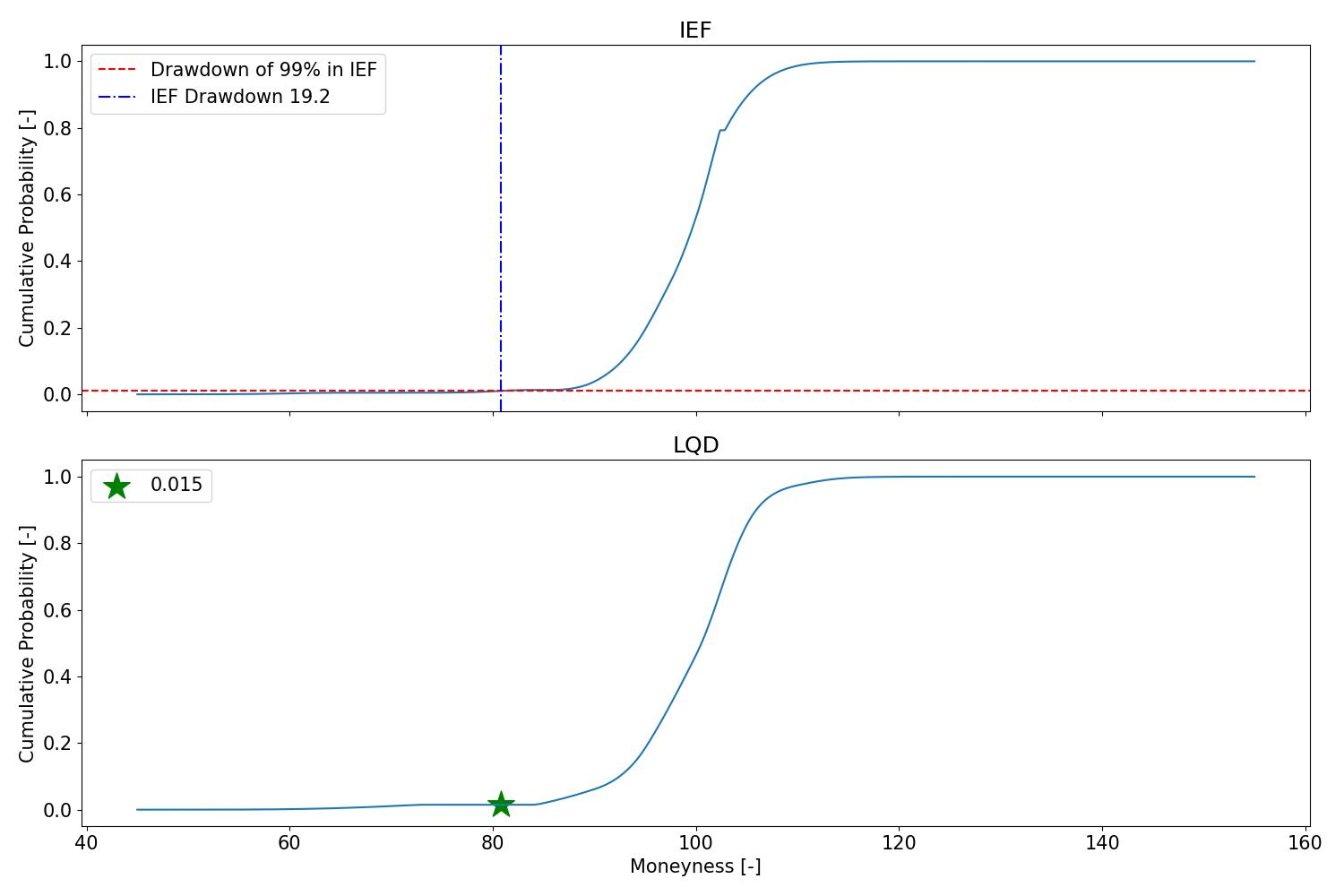}
  \caption{Risk-Neutral CDFs for IEF and LQD on 03-23-2023 Derived from Daily Call Option Implied Volatilities and Illustrating the Duration Removal Process.}
  \label{Duration Removal}
\end{figure}

Rather than applying a formal mathematical process to remove the duration effects from the LQD risk-neutral probabilities, consider the following. If the risk-neutral CDF for IEF reflects a 99\% chance that the ETF will not draw down more than 15\% in the next 3 months, any drawdown for LQD in excess of 15\% is most likely due to credit. Figure \ref{Duration Removal} displays this process pictorially. As a step by step process (with the CDFs already in hand):

\begin{enumerate}
    \item Determine the drawdown associated with a 1\% probability from the IEF CDF.
    \item Determine the point on the LQD CDF that corresponds to the above drawdown.
    \item Numerically integrate the LQD PDF up to this determined drawdown.
\end{enumerate}

\noindent Steps 2 and 3 from the list above represent different ways of looking at the credit-risk factor, with no way of knowing a priori which one will perform better for a hedging signal. For 03-23-2023, Figure \ref{Duration Removal} below shows that the ``credit risk" based on ``Step 2" is $0.50\% = 1.5\% - 1.0\%$.

Performing the above process every day determines the time-series of credit risk, and displayed below. Figure \ref{Credit Risk 1} and \ref{Credit Risk 2} show the two time-series, both with clear peaks during the middle of 2015 and early 2020. The Excess Expected Drawdown (Figure \ref{Credit Risk 2}) looks to be ``cleaner", but both are worth investigating for timing the dynamic hedge.

\begin{figure}[H]
  \centering
  \includegraphics[width=0.80\linewidth]{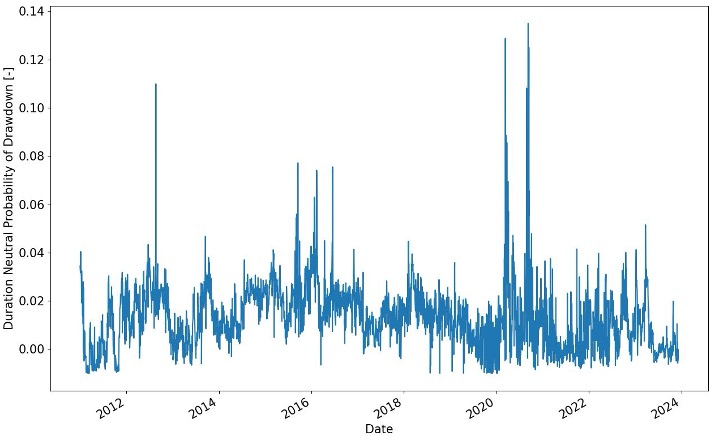}
  \caption{Credit Risk Defined by the Probability of LQD Drawing Down More than IEF in Three Months Time.}
  \label{Credit Risk 1}
\end{figure}

\begin{figure}[H]
  \centering
  \includegraphics[width=0.8\linewidth]{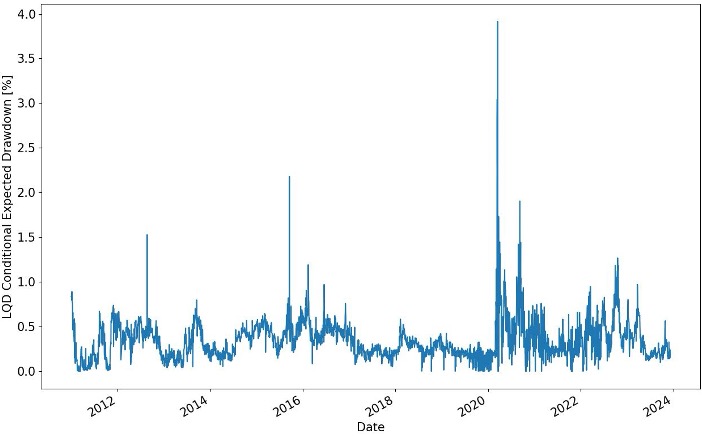}
  \caption{Credit Risk Defined by the Expected LQD Drawdown in Excess of IEF in Three Months Time.}
  \label{Credit Risk 2}
\end{figure}

\section{Liquidity} \label{Liquidity Signal}
Liquidity typically refers to the ``availability" of a financial instrument in the markets and how ``easy" it is for a buyer or seller to enter into a transaction without affecting its market price. Rather than considering the entire corporate bond universe, the current research utilizes TRACE data for the constituents of the 5-year credit default swap index (CDX). The constituents (typically around 130 at any given time) are typically re-balanced every six months, the time series of constituent names also delivered by Bloomberg. 

\subsection{Data Gathering and Cleaning}
The corporate bonds listed by the CDX constituents are typically the most liquid within the investment-grade universe, and the data more reliable. TRACE, being a large database requires a precise and thorough process to parse and acquire the information desired. The method for acquiring and cleaning the data for the IG CDX consituent bonds follows the processes from Dick-Nielsen \cite{dicknielsen2019cleantrace}. More, specifically, the TRACE data cleaning follows a four step process:

\begin{enumerate}
    \item \underline{Standardization of Volume and Other Metrics}: Convert data into numerical values and ensure uniformity in data types, specifically concerning labels.
    \item \underline{Elimination of Same-Day Cancellations}: Identify and expunge records associated with same-day cancellations from the dataset.
    \item \underline{Exclusion of Correction Transactions}: Remove erroneous initial reports and any subsequent transactions marked with a 'C' for 'Correction' in the trade status.
    \item \underline{Removal of Reversal Transactions}: Locate and discard reversal transactions, which were identified by a 'Y' in the trade status and an 'R' as the reversal indicator.
\end{enumerate}

While the above process may seem trivial, it constitutes a significant effort, which should be noted. After cleaning the TRACE data, additional risk measures such as duration and coupon were added for each bond from WRDS monthly bond depository. The spread is then calculated by determining the traded yield to maturity (YTM) as a function of maturity, coupon, and traded price. To calculate the credit spread, we first determined the yield to maturity (YTM) of Treasury securities that match the corporate bonds in terms of maturity and coupon. The spread was then derived by subtracting the Treasury YTM from the traded YTM of the corporate bond. 

\subsection{Signal Construction}

With the time series of constituents in-hand, the liquidity signal herein utilizes individual bond value, duration and spread to construct an average daily metric, i.e.

\begin{equation}\label{L1}
    F_{Liquidity}  = \frac{\sum_{i = 1}^{n} (MktVal_{i,t} \times Duration_{i,t} \times Spread_{i,t} )}{\sum_{i = 1}^{n} MktVal_{i,t}}    
\end{equation}

\noindent where $MktVal_{i,t}, Duration_{i,t}, Spread_{i,t}$ are the market value, duration and spread for each bond in the basket. Note that the summation in Eq. (\ref{L1}) runs over all of the bonds ($i = 1, \ldots ,n$) in the basket up to 3 years after being first included in the IG CDX and computes for each day ($t$). $Duration_{i,t} \times Spread_{i,t}$ is a measure of the excess return volatility of credit securities \cite{bendor2007dts}.

\begin{figure}[H]
  \centering
  \includegraphics[height=0.4\textheight,width=1.0\linewidth]{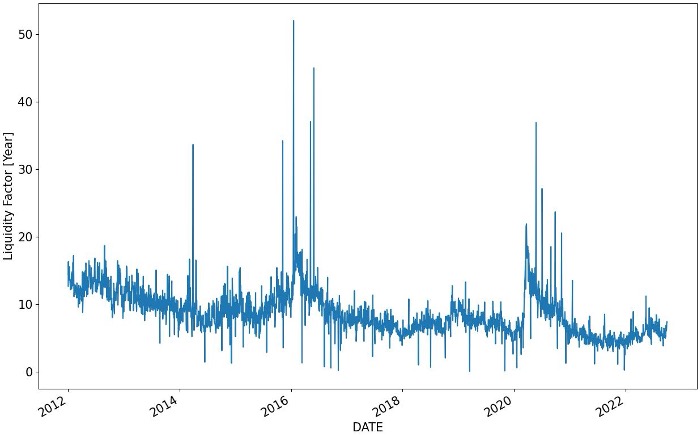}
  \caption{Liquidity Factor Derived from IG CDX Constituent Corporate Bond Data from TRACE.}
  \label{Liquidity Factor}
\end{figure}

Figure \ref{Liquidity Factor} displays the liquidity factor calculated using the above procedure and described mathematically in Eq. (\ref{L1}). The liquidity signal reflects the day-to-day trading activity in the bond market, adjusted for the inherent risk of each bond. By incorporating the bond's duration and credit spread, \(F_{Liquidity} \) captures not just the volume but also the price sensitivity to interest rate changes and the additional yield demanded by investors over the risk-free rate due to credit risk. A higher \(F_{Liquidity} \) value suggests that the market is volatile, with bonds exhibiting higher spreads being traded more frequently, indicating a period where the market is liquidity-demanding. Conversely, a lower \( F_{Liquidity} \) value points to a calmer market environment, where liquidity is more readily supplied. Our aim is to strategically size hedges, adopting a short position in anticipation of periods characterized by high liquidity demand, and covering these shorts when the market transitions to a state of liquidity supply.

\section{Momentum}


Momentum as a cross sectional factor is constructed by sorting assets on realised past returns, and going long a portfolio of top performing assets and short a portfolio of worst performing assets. It is usually formulated as 12m-1m momentum to avoid short-term reversal. Rather than taking the traditional momentum approach, which requires expansive corporate bond data, our formulation seeks to construct a time series momentum signal. We utilise LQD and HYG duration neutral returns as proxies for timing momentum in IG and HY markets, respectively. 

We focus on duration neutral returns as the current dynamic hedge seeks to monetize the widening of IG credit spread, neutralising and abstracting from duration. Appendix \ref{Appendix} details the procedure utilized to remove duration effects from LQD and HYG simple returns, with the resulting return series referred to as duration-neutral $\tilde{R}$.


Define first the n-day duration-neutral cumulative returns for a financial instrument as

\begin{equation}\label{M1}
    \tilde{C}_{n,t} = \frac{\tilde{R}_{t-22}}{\tilde{R}_{t-n-22}} - 1
\end{equation}

where the cumulative return is offset by a business month for 12m-1m momentum.

\noindent With the time-series $\tilde{C}_{n,t}$, define a simple momentum signal utilizing the z-score as

\begin{equation}\label{M2}
    z_{n,t} = \frac{\tilde{C}_{n,t} - \mu_{\tilde{C}_{n,t}}}{\sigma_{\tilde{C}_{n,t}}}
\end{equation}

\noindent where

\begin{equation}\label{M3}
    \begin{gathered}
        \mu_{\Bar{C}_{n,t}} = \frac{1}{n}\sum_{i=1}^n \Bar{C}_{n,t-i} \\
        \text{and} \\
        \sigma_{\Bar{C}_{n,t}} = \sqrt{\frac{\sum_{i=1}^n \big( \Bar{C}_{n,t-i} - \mu_{\Bar{C}_{n,t}} \big)^2} {n - 1}}
    \end{gathered}
\end{equation}

\begin{figure}[H]
  \centering
  \includegraphics[height=0.3\textheight,width=1.0\linewidth]{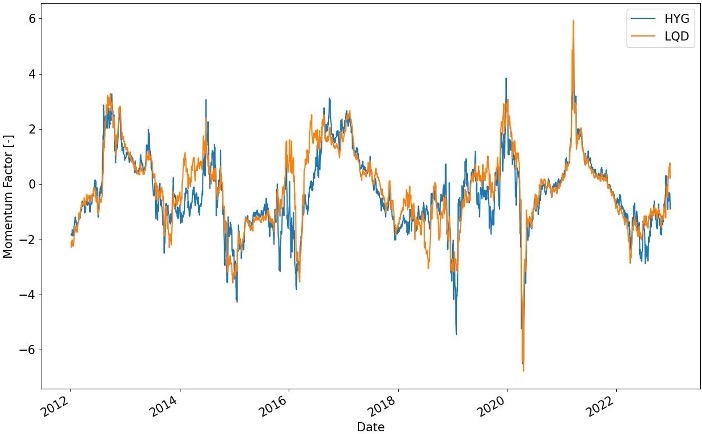}
  \caption{Momentum Factors Derived with LQD and HYG Cumulative Returns and a 252-Day Look-Back Window.}
  \label{Momentum Factor}
\end{figure}

Figure \ref{Momentum Factor} displays the momentum signals constructed for both LQD and HYG utilizing 252 day (i.e. one full trading year) look-back windows. 



Intuitively, the time series momentum signals excluding the last month will be positively correlated to hedge returns in normal times, and negatively correlated in sharp drawdown months and sharp reversal months due to the month-long lag. Momentum signals contain fast information that can help the model distinguish between normal markets with positive momentum and positive hedge returns, onsetting drawdowns with positive past momentum and recent negative hedge returns, and reversals with negative past momentum and recent positive hedge returns.

\section{Signal Content and Orthogonality}

While the literature shows that bond credit risk, liquidity, and momentum signals are important for modelling spread returns of corporate bonds (see the literature review in Chapter \ref{Ch. 2 - Lit Review}), the question of each signal's unique information content is still valid.

\begin{figure}[H]
  \center
  \includegraphics[width=1.0\linewidth]{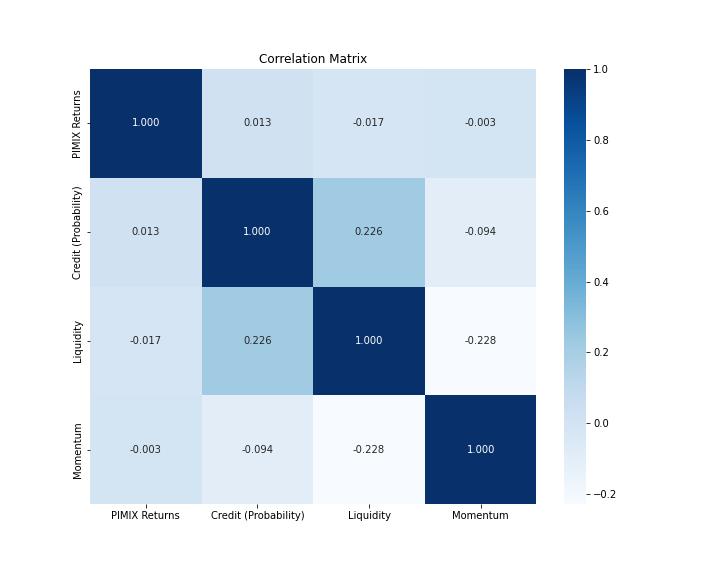}
  \caption{Long-run Correlation Heat Map for PIMIX Returns and the Three Signals Constructed Herein.}
  \label{Signals Correlation}
\end{figure}

Figure \ref{Signals Correlation} displays a correlation map between PIMIX spread returns and the three signals constructed herein (namely, Credit Risk, Liquidity and Momentum). The figure shows that the (long-run) correlations between Momentum, Liquidity and PIMIX returns are somewhat negative while Credit is (slightly) positively correlated. More importantly, the signals themselves have less than 25\% similarity (in Pearson correlation) and all look to have at least a modest amount of orthogonal, additive content.

\noindent Running a simple Vector Auto Regression VAR(0) model

\begin{equation}
\begin{pmatrix} F_{Credit,t} \\ F_{Liquidity,t} \\ F_{Momentum,t} \\ \end{pmatrix} = \alpha + 
\begin{pmatrix} F_{Liquidity,t} & F_{Momentum,t} \\ F_{Credit,t} & F_{Momentum,t} \\ F_{Credit,t} & F_{Liquidity,t} \\ \end{pmatrix}
\begin{pmatrix} \beta_{1} & \beta_{2} \end{pmatrix} +
\begin{pmatrix} u_{1} \\ u_{2} \\ u_{3} \end{pmatrix}
\end{equation}


\noindent also helps determine whether or not the signals are orthogonal. Running the system through equation by equation OLS (consistent for stationary series) delivers low adjusted R-squared $\leq 0.05$, showing that the signals are indeed (linearly) unrelated. 

\clearpage
\chapter{Hedging Strategy Optimization} \label{Ch. 4 - Model}

\vspace{-3mm}
The following section addresses two methods for combining the three signals derived in Chapter \ref{Ch. 3 - Signals} and a methodology for when to establish a short position in LQD (or HYG, or both) to hedge a bond mutual-fund. Express the simple (duration neutral) returns of a hedged portfolio as

\begin{equation} \label{strat1}
    \tilde{R}_{hedged\ portfolio,t} = \tilde{R}_{mutual\ fund,t} + \sum_{i=1}^n W_{i,t} \tilde{R}_{hedge_{i},t}
\end{equation}

\section{Simple Two-Step Ordinary Least Squares}

The construction of our credit market signals facilitates a trading strategy that capitalizes on explanatory power these signals provide on the hedge instrument returns.

Firstly, consider a strategy which attempts to explain the duration-neutral returns ($\hat{R}$) of the hedge instrument (either LQD or HYG) by a weighted linear combination of the three signals (credit risk, liquidity and momentum). The simple model can be expressed as

\begin{equation}\label{OLS1}
    \tilde{R}_{hedge_{i},t+1} = \alpha + \beta_{Credit} F_{Credit,t} + \beta_{Liquidity} F_{Liquidity,t} + \beta_{Momentum} F_{Momentum,t}
\end{equation} 

\noindent where $\tilde{R}_{hedge_{i}}$ is a column vector of the hedge's duration neutral returns, and $F_{Credit}, \text{ } F_{Liquidity}, \text{ } \\F_{Momentum}$ are column vectors of one day lagged credit, liquidity and momentum signal observations. Rolling Ordinary Least Squares (OLS) determines the coefficients in Eq. (\ref{OLS1}), the process itself being independent of the regression window chosen. An appropriate regression lookback window would balance recent history with sufficient data samples, and we therefore focus our analysis on a quarter (90 days).


Ideally, the hedge activates when it's profitable to short the hedging instrument (i.e. a short position would earn a positive next-day return). Consider now two criteria, which in conjunction determine when to activate the hedge:

\begin{center}
    \item $\hat{R}_{hedge_{i}, t+1} < 0$
    \item $p(F_{statistic}) \leq \gamma_{OLS}$
\end{center}

\noindent The first criteria from the list above ensures that the hedge turns on when the next-day forecasted return is negative. Not only does the predicted return need to be negative, but the p-value of the model's F-statistic (which relates the weighted ratio of the explained to unexplained variance in the regression) must be less than $\gamma_{OLS}$, which reflects the modelled relationship's statistical significance. As with the regression window, there is likely an optimal $\gamma_{OLS}$ for the combination of hedge instrument and the three signals. \\

An indicator variable succinctly summarises the two criteria from above as
\vspace{-2.5mm}
\begin{equation}\label{OLS2}
    \mathbbm{1}_{hedge_{i}, OLS, t+1} =
    \begin{cases}
    1, & \text{if } \hat{R}_{hedge_{i}, t+1} < 0 \text{ and } p(F_{statistic}) \leq \gamma_{OLS} \\
    0, & \text{otherwise}
    \end{cases}
\end{equation}

The process above determines when to apply the hedge, but not the weight on the short hedge position per unit of the invested portfolio. A second step regression decides the hedge's weight, setting it equal to the empirical $\beta$ computed through a regression of the bond portfolio's returns on the hedge instrument's returns.
\begin{equation} \label{OLS3}
\bold{\tilde{R}_{mutual\ fund}} = \kappa + \beta_{hedge_{i}}\bold{ \tilde{R}_{hedge_{i}}}
\end{equation}

\noindent Finally, the live hedge position is the product of the $\beta_{hedge}$ weight and the activation indicator from Eq. (\ref{OLS2}).

\vspace{-2.5mm}
\begin{equation}\label{OLS4}
    W_{hedge_{i},t+1} = -\beta_{hedge_{i}} \cdot \mathbbm{1}_{hedge_{i}, OLS,t+1}
\end{equation}

Substituting Eqs. (\ref{OLS1} - \ref{OLS4}) into Eq. (\ref{strat1}) gives an expression for the hedged portfolio value in terms of the regression coefficients,

\begin{equation}\label{OLS5}
    \begin{aligned}
        \bold{\tilde{R}_{hedged\ portfolio}} &= \bold{\tilde{R}_{mutual\ fund}} + \sum_{i=1}^n \bold{(-\beta_{hedge_{i}}} \cdot \bold{\mathbbm{1}_{hedge_{i}, OLS})} \cdot \bold{\tilde{R}_{hedge_{i}}}
    \end{aligned}
\end{equation}

\noindent Empirical observations (displayed later) reveal that the F-statistic and R-squared scores of the first step regression increase during hedge instrument drawdowns, indicating our signals have strong predictive power in past drawdown periods and are useful for activating hedges.


\section{Maximising Orthogonal Correlation}

Linear regression provides an adequate ``baseline" model for hedge timing and for setting hedge weights. Another method, Canonical Correlation Analysis (CCA) is capable of optimize and solve for hedge instruments weights in a single step. The CCA algorithm takes as inputs two time series and aims to assign weights to each to maximize the correlation between the two constructed variables. In mathematical terms, given a matrix $\bold{Y}$ of independent variable column vectors and matrix $\bold{X}$ of independent feature column vectors:

\begin{equation}\label{CCA1}
  \bold{X}={(\bold{x_1},\ldots,\ \bold{x_k})}, \quad \bold{Y}={(\bold{y_1},\ldots,\ \bold{y_m})}
\end{equation}



\noindent canonical correlation analysis seeks weight vectors $\boldsymbol{a}\ (\boldsymbol{a}\in \mathbb{R}^k)$ and $\boldsymbol{b}\ (\boldsymbol{b}\in \mathbb{R}^m)$ to maximize correlation between $\boldsymbol{Xa}$ and $\boldsymbol{Yb}$. This is equivalent to selecting the first 'principal component' (linear combination of vectors) in the dependent and feature space, so as to maximise between features and returns.

\subsubsection{Hedging with a Single Hedge Instrument}

For a hedged portfolio comprised of the mutual fund and a single hedge asset, $\bold{Y} \in \mathbb{R}^{n \times 2}$. The three signals (liquidity, momentum, and risk) comprise $\bold{X} \in \mathbb{R}^{n \times 3}$.


\begin{equation}\label{CCA2}
    \begin{gathered}
        \bold{Y}={(\bold{\tilde{R}_{mutual\ fund}},\bold{\tilde{R}_{hedge})}} \\
        \bold{X}={(\bold{F_{Liquidity}},\bold{F_{Momentum}},\ \bold{F_{Credit})}}  
    \end{gathered}
\end{equation}


The first pair of canonical weights represent the linear combination of factors that maximise correlation to the hedged portfolio's duration neutral returns. CCA can be formulated as a maximisation of the following objective function:


\begin{equation}\label{CCA3}
    (\boldsymbol{a}^*, \boldsymbol{b}^*) = \max_{(\boldsymbol{a}, \boldsymbol{b})} \text{corr}(\boldsymbol{Xa}, \boldsymbol{Yb})    
\end{equation}

\noindent where

\begin{equation}\label{CCA4}
    \boldsymbol{a}^\ast=(w_{mutual\ fund}^\ast,\ W_{hedge_{i}}^\ast)^T=(1, W^*_{hedge_{i}})^T
\end{equation}

\begin{equation}\label{CCA5}
  \boldsymbol{b}^\ast=(w_{Liquidity}^\ast,\ w_{Momentum}^\ast, \ w_{Credit}^\ast) ^T
\end{equation}

One can easily show that any proportional change to $\boldsymbol{a}^*$ or $\boldsymbol{b}^*$ still guarantees the optimal solution. In the current analysis, the investor always holds 1 unit of the mutual fund $w_{\text{mutual fund}}^* = 1$, so the hedge's weight is normalised per unit of mutual fund, $w^* = \frac{w_{\text{hedge}}^*}{w_{\text{mutual\ fund}}^*}$.

As with the 2-step linear regression from the previous section, the CCA provides an optimal weights for the hedged position, but does not give any information as to when to activate the hedge. A second-step regression follows the CCA, modelling a linear relationship between the CCA hedged portfolio's $\boldsymbol{Yb} \in \mathbb{R}^n$ returns and the CCA optimal $\boldsymbol{Xa} \in \mathbb{R}^n$ joint signal component.


\begin{equation}\label{CCA6}
    \tilde{R}_{hedged\ portfolio, t+1}^\ast = \omega + \beta (\boldsymbol{Xa})_{t} + \epsilon_{t+1}
\end{equation}

Note that throughout this paper, and as in Eq. (\ref{CCA6}), we are using yesterday's signals to predict tomorrow's hedged return, preventing data leakage and avoiding look-ahead bias. \\

As before, the following two criteria determine whether or not to initiate the hedge:

\begin{center}
    $W_{hedge_{i},t}^\ast < 0$ \\
    $\hat{z}_{CCA,t+1} > \gamma_{CCA_{upper}}$
\end{center}

\noindent where $\hat{z}_{CCA,t+1}$ is the z-score of the next day-forecasted return delivered by Eq. (\ref{CCA6}) and calculated similar to Eqs. (\ref{M2} \& \ref{M3}). The first condition requires that the optimal weight for the hedge instrument be negative (i.e. a short position) while the second ensures hedge initiation occurs only when the hedged portfolio next-day forecasted returns are significantly above average (i.e. the prediction's z-score is above the $\gamma_{CCA}$ threshold).

In simpler models, trading in and out of the hedge was unconstrained, assuming abundant volume was available to enter and exit the hedge. To make the CCA hedging framework more applicable to hedging larger funds with billions in assets, we introduce an additional rule to time our exit. This rule is very additive, lowering hedge turnover, and increasing drawdown resilience through keeping the hedge on for longer. 

To take off the hedge, the regression must predict a forecasted hedged return that is significantly below average and below the (often negative) $\gamma_{CCA_{lower}}$ threshold.

\begin{center}
$\hat{z}_{CCA,t+1} < \gamma_{CCA_{lower}}$
\end{center}

This indicates that duration neutral hedge returns are starting to rebound and mean revert, and helps us make a timely exit post drawdown.\\

The indicator variable that summarizes these conditions succinctly is

\begin{equation}\label{CCA7}
    \mathbbm{1}_{hedge_{i}, CCA,t+1} =
    \begin{cases}
    1, & \text{if } W_{hedge_{i},t}^\ast < 0 \text{ and } \hat{z}_{CCA,t+1} >\gamma_{CCA_{upper}} \\
    1, & \text{if } \mathbbm{1}_{hedge_{i}, CCA,t} = 1 \text{ and } \hat{z}_{CCA,t+1} \geq \gamma_{CCA_{lower}} \\
    0, & \text{if } \mathbbm{1}_{hedge_{i}, CCA,t} = 1 \text{ and } \hat{z}_{CCA,t+1} < \gamma_{CCA_{lower}} \\
    0, & \text{otherwise}
    \end{cases}
\end{equation}


\noindent Note that $\gamma_{CCA_{upper}}$ and $\gamma_{CCA_{lower}}$ are the 
``on" and ``off" thresholds for the hedge timing and do not need to be symmetric. Finally, to ensure this strategy is a true ``hedge" instead of a tactical net short, we cap $|W_{hedge_{i,t}}| \leq 1$ such that its weight can never exceed that of the mutual fund. We also apply volatility scaling to size hedge weights more effectively for funds with different active volatility targets, setting $|W_{hedge_{i,t}}| \leq \min(\sigma_{fund}/\sigma_{hedge_{i}},1)$, where $\sigma$ denotes last year's duration neutral return volatility.

\subsubsection{Hedging with Multiple Hedge Instruments}
When dealing with multiple hedge instruments, ensuring the CCA learns negative weights for all hedge instruments in the canonical covariate $\boldsymbol{Xa} \in \mathbb{R}^n$, as in $\boldsymbol{a<0}$, is challenging. To overcome this, Principal Component Analysis is applied on the hedge returns, creating a single, representative hedge instrument. Since hedges are typically positive correlated, the first PC closely resembles a simple weighted average of all instrument returns. Therefore $W_{hedges,t}$ takes the interpretation of an optimal weight to a synthetic hedge, which is distributed into individual hedges through $W_{hedges,t} \cdot PC_{1}$ when trading into a hedge.

\subsection{Adjusting for Market Impact and Transaction Costs}

The above framework considers the mathematics of when to turn on the hedge, but doesn't account for trading slippage, volume informed realistic position taking, or short hedge position funding costs. To address these, unit portfolio returns rom  Eq. (\ref{strat1}) are modified to include transaction costs computed using live bid and ask spreads (through $P_{ask} \text{ and } P_{bid}$), and funding costs (denoted $f_{bps}$).

\begin{equation} \label{Costs1}
\begin{aligned}
    \tilde{R}_{hedged\ portfolio, t} = \tilde{R}_{mutual\ fund, t} + \sum_{i=1}^n W_{i, t} \bigg[\tilde{R}_{hedge_{i},t} &- 
    \mathbbm{1}_{trade_{i},t}\bigg(\frac{P_{ask,i,t} - P_{bid,i,t}}{P_{ask,i,t} + P_{bid,i,t}}\bigg) \\
    & - \mathbbm{1}_{hedge_{i},t}\bigg( \frac{f_{bps,i,t}}{1000}\bigg)\bigg]
\end{aligned}
\end{equation}


\noindent As a simplification, note that Eq. \ref{Costs1} considers the best bid and ask symmetric about the midprice. In addition, the indicator variable $\mathbbm{1}_{trade_{i}} = 1$ only on days where hedge$_{i}$ is actively sold short or bought to establish or exit the position. 

Finally, to incorporate some degree of realism for position taking and market effects, the current framework imposes a volume based restriction on the hedge activation. Consider an n-day simple moving average of the hedge instrument's historical daily trading volume ($V$),

\begin{equation}\label{Costs2}
    V_{t,d}' = \frac{1}{d}\sum_{j=0}^{d-1} V_{t-j} 
\end{equation}

A volume ($V$) based restriction for hedge activation and deactivation limits the daily volume traded to 10\% of the 252-day (i.e., one trading year) simple moving average of the hedge instrument's daily trading volume. Mathematically,

\begin{equation}\label{Costs3}
    V_{hedge, t} \leq \textbf{min}\bigg(\frac{V_{t, 252}'}{10} \text{, } V_t \bigg)
\end{equation}

\noindent Thus, if the volume of the hedge instrument suggested by the optimal weight exceeds 10\% of the moving average on any given day, the hedge gradually accumulates (or removes) over multiple consecutive trading days, minimising market impact until sufficient liquidity appears to complete full entry or exit into and out of the hedge.
\clearpage
5Se\chapter{Performance of the Dynamic Hedge for a Bond Portfolio} \label{Ch. 5 - Results}

As of the end of January, 2024, Pimco’s PIMIX fund represented $\approx 126$bln total net assets, nearly one percent of the total notional outstanding value in the IG bond market. Chapter 1 discussed briefly the extension in the IG bond market since the GFC, increasing from roughly 5 trillion USD to ~10 trillion as late as 2019 (presumably even larger now in 2024 after large net issuances post COVID quantitative easing). The following chapter considers 
the hedge's efficacy in protecting a generic bond fund, here PIMIX's spread returns, at first without considering transaction, funding costs, or market impact of putting on or taking off the short positions. 

After showing the efficacy of the strategy without realistic market frictions, the next section includes these frictions (trading costs, funding costs and volume-based position taking limits) and for bond portfolios with total net assets of 100 million, 1 billion and 10 billion, representing small, medium, and large credit portfolios. The research also considers different combinations of Credit, Liquidity and Momentum signals, and multiple hedge instruments across credit ETFs (LQD, HYG, JNK) and credit CDXs (IG 5YR, HY 5YR). The priors we set are that all signals together work best, and that shorting ETFs can better capture downside convexity compared to CDXs and are therefore better hedges. 


\section{Hedging PIMIX with a Single Instrument}\label{sec:Single Hedge}

\subsection{Credit ETF: LQD}
Consider first a generic bond fund (represented by PIMIX returns) and as a proof-of-concept, hedging the fund with LQD only. The CCA method times when to put on and take off the hedge, as described in Chapter \ref{Ch. 4 - Model} and with the total back-test extending from 01/2013 – 09/2022 due to lack of duration information for the hedge assets (LQD, HYG, JNK) prior to and after those dates\footnote{As discussed in the Introduction, this timeframe is realistic for shorting corporate bond ETFs, as HYG was still in its infancy, initiated in 2007.}. The following section presents results for the CCA informed hedge, as it outperforms the 2-Step OLS method when comparing hedged portfolio risk characteristics including the Sortino Ratio.


Ideal hedges should have relatively cheap negative carry and large payoffs during drawdowns, effectively limiting portfolio downside. Hedges add value in two ways: short-term payoffs that cancel out drawdowns neutralising risk, and long-term benefits from retaining fund value allowing for larger compounding gains. In that regard, hedges should at least cut down risk and drawdown, and potentially be net profitable from higher expected return over the longer run. One metric that encapsulates all of the above is the Sortino Ratio, expressing average annual (duration neutral) return of a portfolio, as normalised by downside standard deviation of (duration neutral) returns.


\begin{equation}\label{R1}
    \text{Sortino Ratio} = \frac{\mu_{hedged}}{\sigma_{hedged,d}}
\end{equation}


\noindent and where $\mu_{hedged}$ represents average annualised hedged portfolio returns, and $\sigma_d$ represents the annualised standard deviation of the negative (or drawdown) returns. The Sortino ratio represents how ``good" portfolio performance is, expressed as a function of ``bad" times denoted by downside risk during drawdowns.   

The three main parameters that inform the model are: lookback window for training, and z-score predicted hedged return thresholds ($\gamma_{CCA, lower}, \gamma_{CCA, upper}$) for turning the hedge on and off. The analysis herein displays a wide parameter range grid-search that sequentially searches over the parameter space and converges to a local optimum. This method is deployed as it displays model robustness and performance over ranging model parameter choices, rather than seeking to find a global model optimum.



\begin{figure}[H]
  \center
  \includegraphics[width=0.85\linewidth]{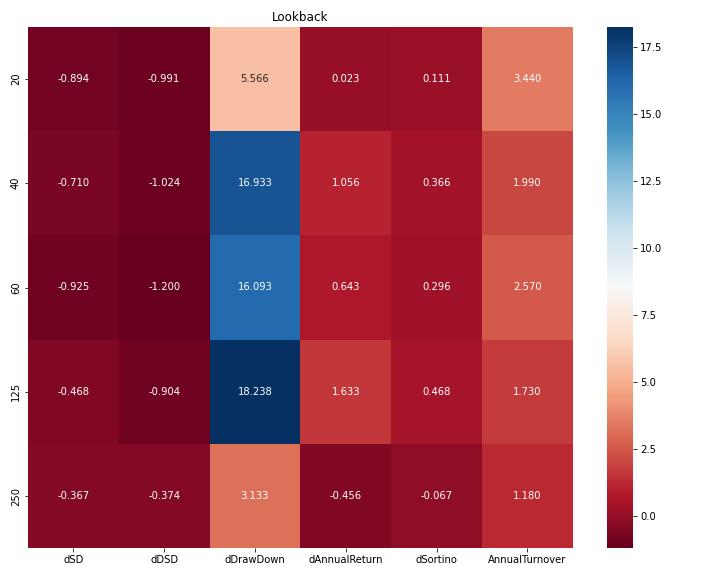}
  \caption{Heat-map Displaying Hedge Performance for Different CCA Lookback Training Windows.}
  \label{optimize 1}
\end{figure}

\begin{figure}[H]
  \center
  \includegraphics[width=0.85\linewidth]{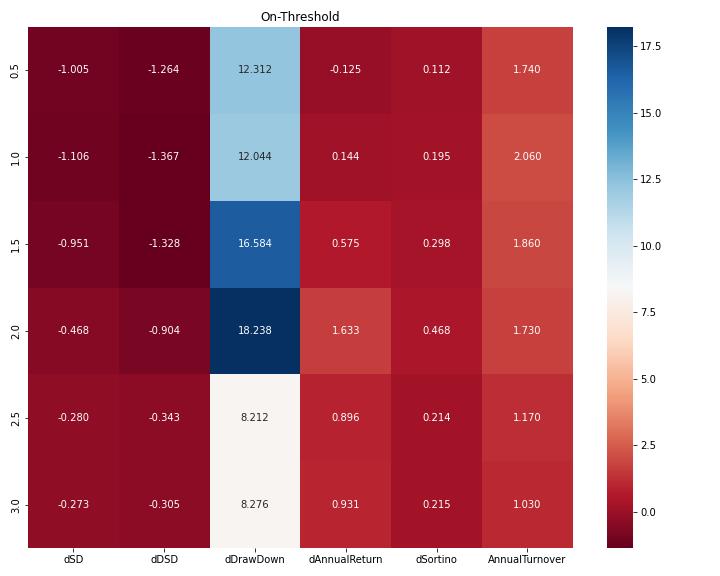}
  \caption{Heat-map Displaying Hedge Performance for Different CCA On-Thresholds.}
  \label{optimize 2}
\end{figure}

\begin{figure}[H]
  \center
  \includegraphics[width=0.85\linewidth]{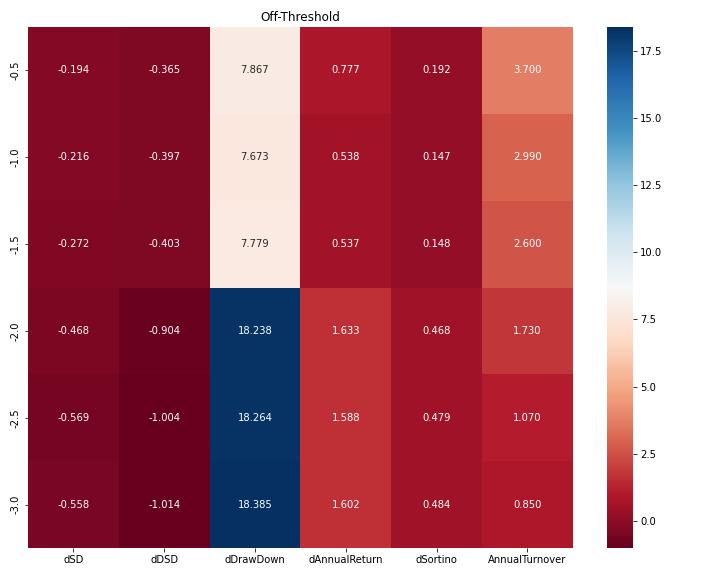}
  \caption{Heat-map Displaying Hedge Performance for Different CCA Off-Thresholds.}
  \label{optimize 3}
\end{figure}

Figures \ref{optimize 1} - \ref{optimize 3} show the step-by-step optimization process, starting first with the lookback window, moving on to the on-threshold and then finally to the off-threshold. Each heat-map compares (from left to right) the changes in standard deviation, downside standard deviation, maximum drawdown, annual return, Sortino ratio and turnover between the hedged portfolio. Differences are expressed as hedged portfolio metrics minus baseline portfolio metrics: $dmetric = metric_{hedge}-metric_{baseline}$. An ideal hedged portfolio would decrease standard deviation, decrease downside standard deviation, limit large drawdowns (dradown$_{hedge}=-10\%$, drawdown$_{baseline}=-25\% \implies$  $ddrawdown=+15\%$), increase annual return, increase the Sortino ratio, and have low annual turnover. All these objectives seem to (simultaneously) be achieved by the CCA hedge in the following gridsearches.

\clearpage

Referring first to Figure \ref{optimize 1}, the grid search utilizes 20, 40, 60, 125 and 250 business day (match 1, 2, 3, 6 and 12 month) lookback windows for the local optimization. Exploring the heat-map, the hedged portfolio performs better than the baseline (on average) for all lookback windows except for 250 days. Midsized lookback windows strike a good balance between recent history and sufficient training data, reducing standard deviations, and largely avoiding drawdowns, and positively contributing to returns - thereby boosting Sortino. The short-term 20bd window has little data to train on and high turnover, whereas the longest window of 250bd is slower and rarely turns on the hedge. When optimizing based on Sortino Ratio, the locally optimal lookback is 125 trading days (approximately 6 months).


\clearpage

With the optimal rolling training-window determined, the next step optimization (Figure \ref{optimize 2}) shows that the locally optimal on-threshold z-score ($\gamma_{CCA,upper}$) of the predicted next-day returns is 2.0. A lower on-threshold should result in the hedge activating more often, and indeed, the annual turnover is inversely proportional to $\gamma_{CCA,upper}$. The Sortino ratio shows a somewhat parabolic relationship with the on-threshold, increasing up to a point ($\gamma_{CCA,upper} = 2.0$) and after which exhibiting a sharp drop off. 

\clearpage

Finally, and with the first two parameters tuned, the locally optimal off-threshold $\gamma_{CCA,lower} = -3.0$. Interestingly, optimal $\gamma_{CCA,lower}$ could realistically be either -2.0, -2.5 or -3.0. The annual turnover decreases with decreasing (in magnitude) off-threshold, thus the -2.0 threshold may actually be better when considering market frictions and realistic trading conditions.

\begin{figure}
    \caption{$\bold{CCA \ Portfolio}$ $\bold{PIMIX}$, $\bold{LQD \ Hedge}$, $\bold{Liquidity \ Signal}$, $\bold{No \ Costs}$.}

    \centering
    \begin{subfigure}{\linewidth}
        \centering 
        \includegraphics[width=\linewidth]{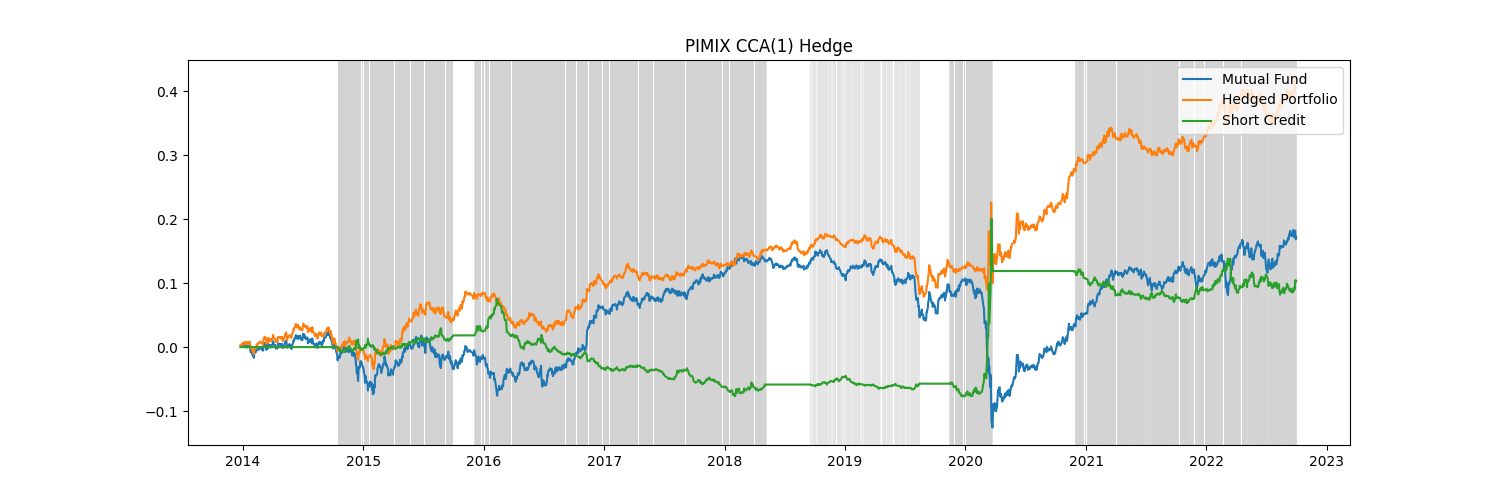}
        \caption{Cumulative Returns}
        \label{LQD_Liq_returns}
    \end{subfigure}

    \vspace{0.5cm} 
    
    \begin{subfigure}{\linewidth}
        \centering 
        \includegraphics[width=\linewidth]{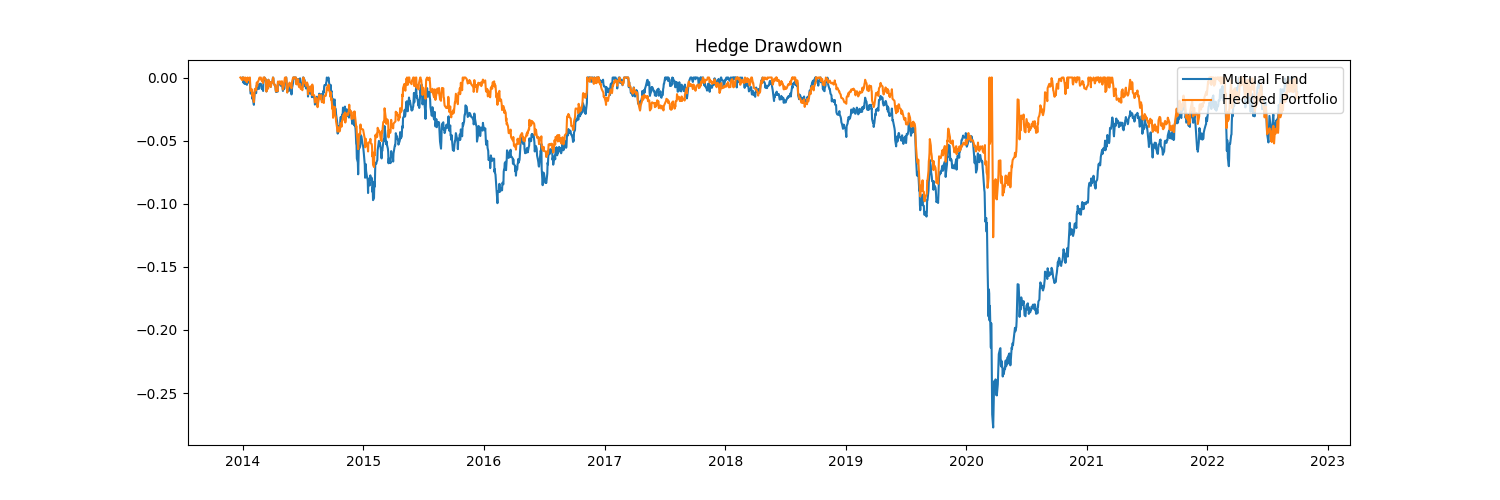}
        \caption{Drawdowns}
        \label{LQD_Liq_drawdowns}
    \end{subfigure}

    \vspace{0.5cm} 
    
    \begin{subfigure}{\linewidth}
        \centering 
        \includegraphics[width=\linewidth]{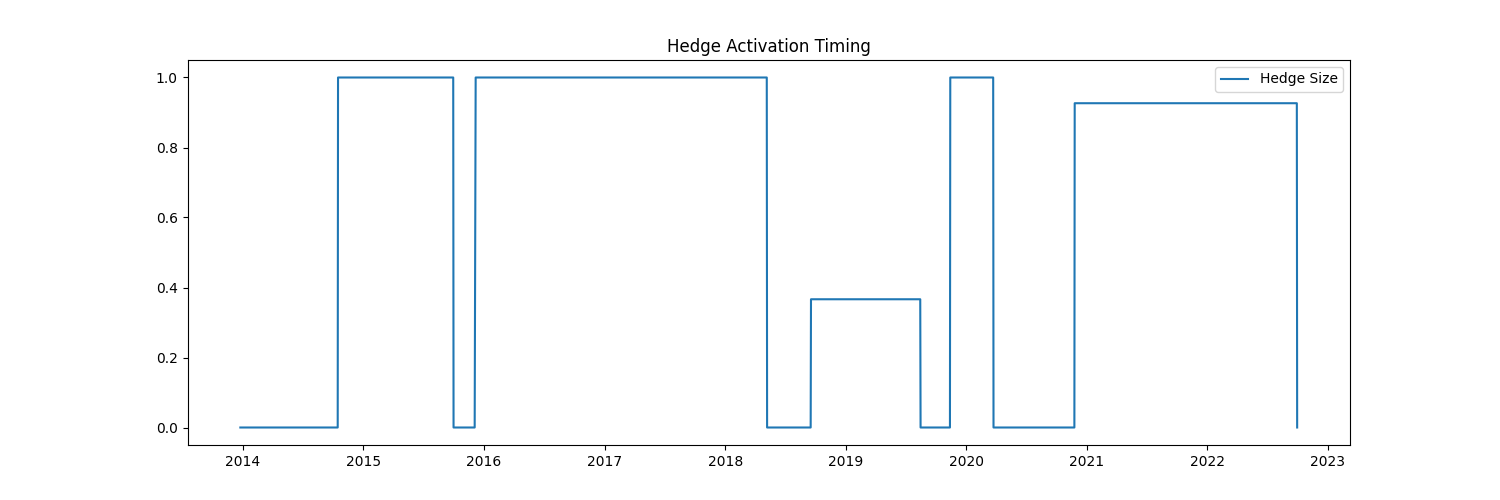}
        \caption{Hedge Weights and Activations}
        \label{LQD_Liq_activations}
    \end{subfigure}
    \label{LQD_Liquidity}
\end{figure}

Figure \ref{LQD_Liq_returns}-\ref{LQD_Liq_activations} displays the cumulative returns and percent drawdowns of the portfolios (hedged and unhedged) as well as the timing of the dynamic hedge during the backtest period. Without trading frictions, the hedge portfolio performance exceeds that of the unhedged, avoiding large drawdowns during several market downturns. The majority of the outperformance comes from avoiding the large drawdown during the 2020 COVID pandemic, however. The average Sortino ratio for the hedged portfolio improves more than approximately 50\% as compared to the baseline. Note that this particular hedge parametrisation is fairly passive and on for a large portion of the backtest, making it viable only on liquid low funding cost instruments like LQD. We investigate funding cost liability in the next section.

Table \ref{LQD Tuned Parameters} shows the tuned parameters as well as the increase in hedged portfolio Sortino Ratio for the portfolio hedged with LQD and using different combinations of the signals (Credit Risk, Liquidity and Momentum) to time the hedge execution. Note that when used individually, Credit Risk and Liquidity perform roughly the same while the hedge timed with Momentum only is less effective. As suggested by the relative orthogonality of the signals, combining different signals generally increases hedged portfolio performance. Encouragingly, the best hedged portfolio performance occurs when the CCA considers all three signals.

\begin{table}[H]
    \centering
    \caption{Tuned Parameters and Increase in Sortino Ratio for the Hedged Portfolio  for Different Signals Considered (Liquidity, Momentum \& Credit: LSG, MSG \& CSG).}
    \begin{tabular}{>{\centering\arraybackslash}m{1.5in}>{\centering\arraybackslash}m{0.75in}>{\centering\arraybackslash}m{0.75in}>{\centering\arraybackslash}m{0.75in}>{\centering\arraybackslash}m{1.25in}}
    \toprule
    Signals & Lookback [Days] & $\gamma_{CCA,lower}$ [std] & $\gamma_{CCA,upper}$ [std] & $\Delta \text{Sortino Ratio}_{252}$ $\text{ [std}^{-1}]$\\
    \midrule
    CSG & 250 & -1.0 & 2.5 & 0.467 \\
    LSG & 125 & -3.0 & 2.0 & 0.468 \\
    MSG & 20 & -0.5 & 3.0 & 0.099 \\
    LSG, MSG & 40 & -1.5 & 2.0 & 0.451 \\
    LSG, CSG & 125 & -2.0 & 2.5 & 0.569 \\
    MSG, CSG & 20 & -3.0 & 3.0 & 0.063 \\
    \bottomrule
    LSG, MSG, CSG & 40 & -1.50 & 2.50 & 0.787 \\
    \bottomrule
    \end{tabular}
    \label{LQD Tuned Parameters}
\end{table}

\begin{figure}
    \caption{$\bold{CCA \ Hedged}$ $\bold{PIMIX}$, $\bold{LQD \ Hedge}$, $\bold{All \ Signals}$, $\bold{No \ Costs}$.}
    \centering
    \begin{subfigure}{\linewidth}
        \centering 
        \includegraphics[width=\linewidth]{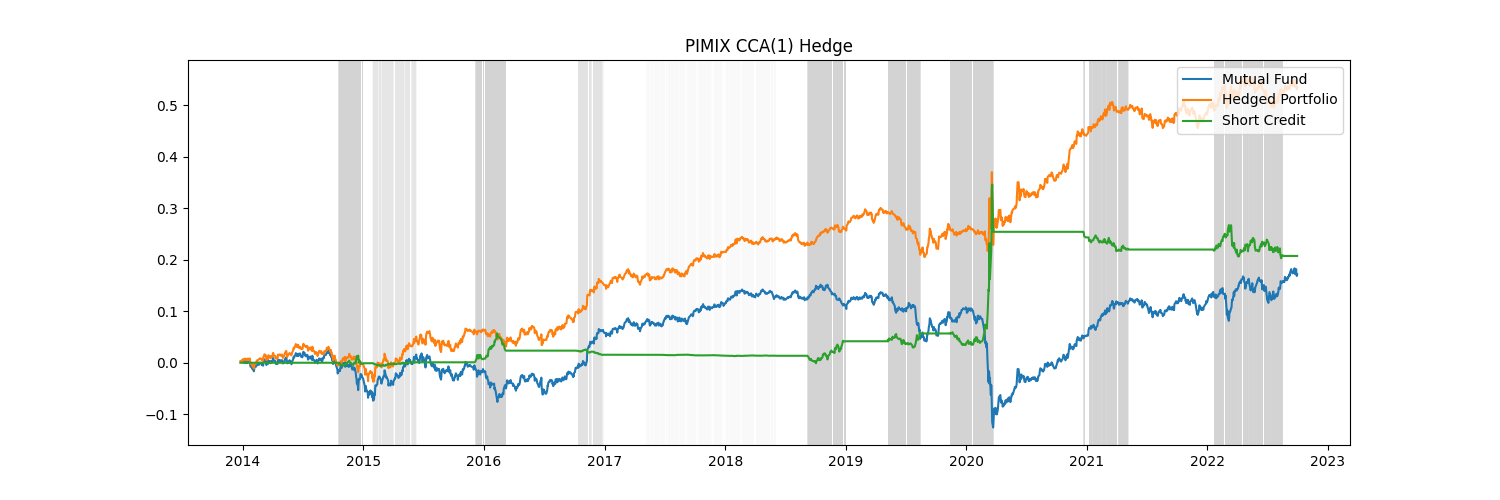}
        \caption{Cumulative Returns}
        \label{LQD_combined_returns}
    \end{subfigure}

    \vspace{0.5cm} 
    
    \begin{subfigure}{\linewidth}
        \centering 
        \includegraphics[width=\linewidth]{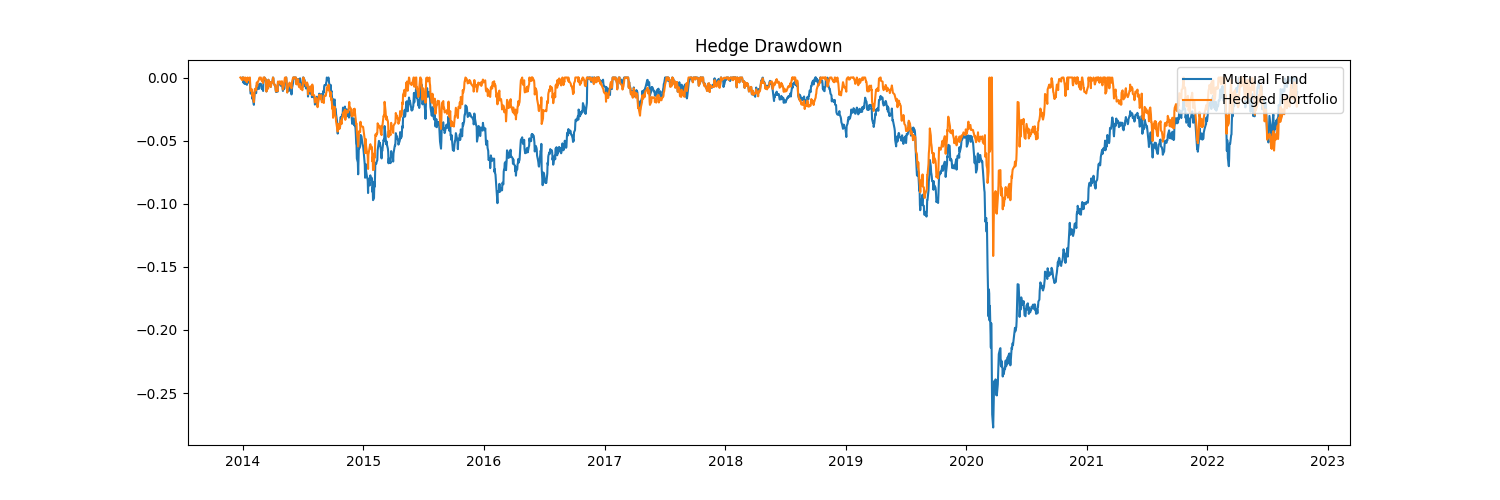}
        \caption{Drawdowns}
        \label{LQD_combined_drawdowns}
    \end{subfigure}

    \vspace{0.5cm} 
    
    \begin{subfigure}{\linewidth}
        \centering 
        \includegraphics[width=\linewidth]{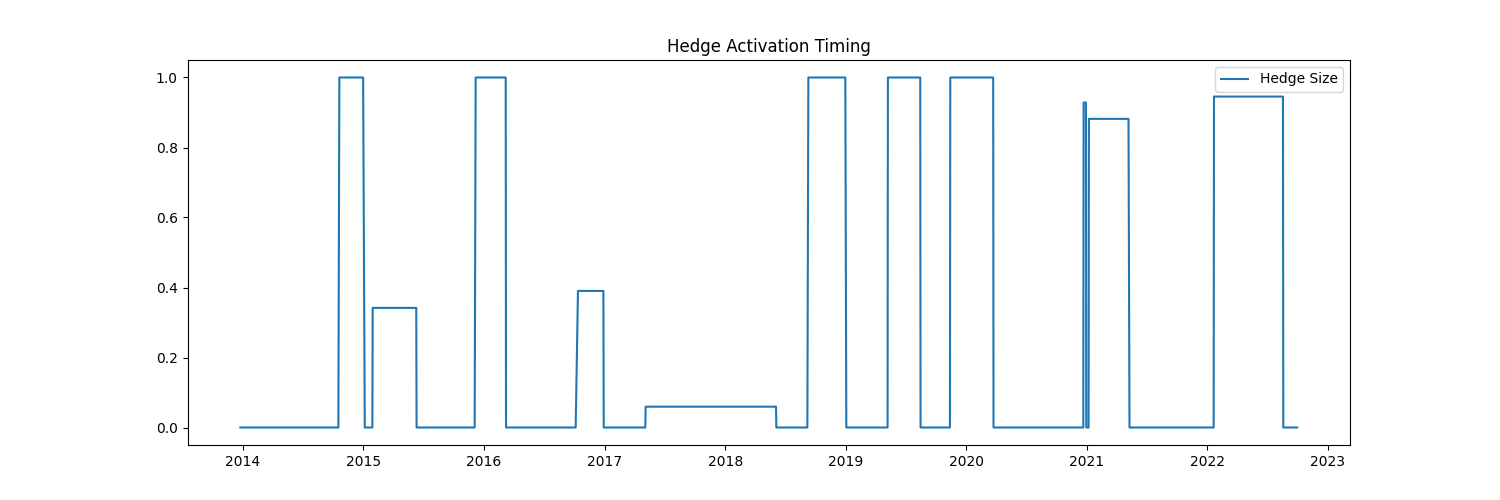}
        \caption{Hedge Weights and Activations}
        \label{LQD_combined_activations}
    \end{subfigure}
    
    \label{LQD_All_no_costs}
\end{figure}

Figure \ref{LQD_combined_returns}-\ref{LQD_combined_activations} shows again the cumulative returns, drawdown and hedge weights for the PIMIX portfolio hedged with LQD only, but this time, considering all three signals. As displayed in Table \ref{LQD Tuned Parameters}, the hedged portfolio has superior performance, avoiding (at least some) drawdowns in 2015, 2017, 2018, 2019 and 2020. To pick up these different market events, the hedge activates much more regularly than with the Liquidity only signal. These results are very encouraging, but again, note that these results assume that the portfolio manager can hedge even the entire notional value in a single day and that the market has no frictions (transaction or funding costs).

\clearpage

It is very additive to visualise how the CCA algorithm learns to effectively hedge a portfolio. In the above backtest, where PIMIX is hedged using LQD and hedge entry/exit timing is informed through CCA on all signals (Credit, Liquidity, Momentum), plotting the achieved canonical correlation reveals that correlation between the joint signals and hedged return spikes during credit market drawdowns - meaning signals are highly informative on downside protection. CCA learns to enter when correlation starts to spike, and either exit near the peak or slightly later, when correlations unwind to previous levels.

\begin{figure}[H]
  \center 
  \includegraphics[width=1.0\linewidth]{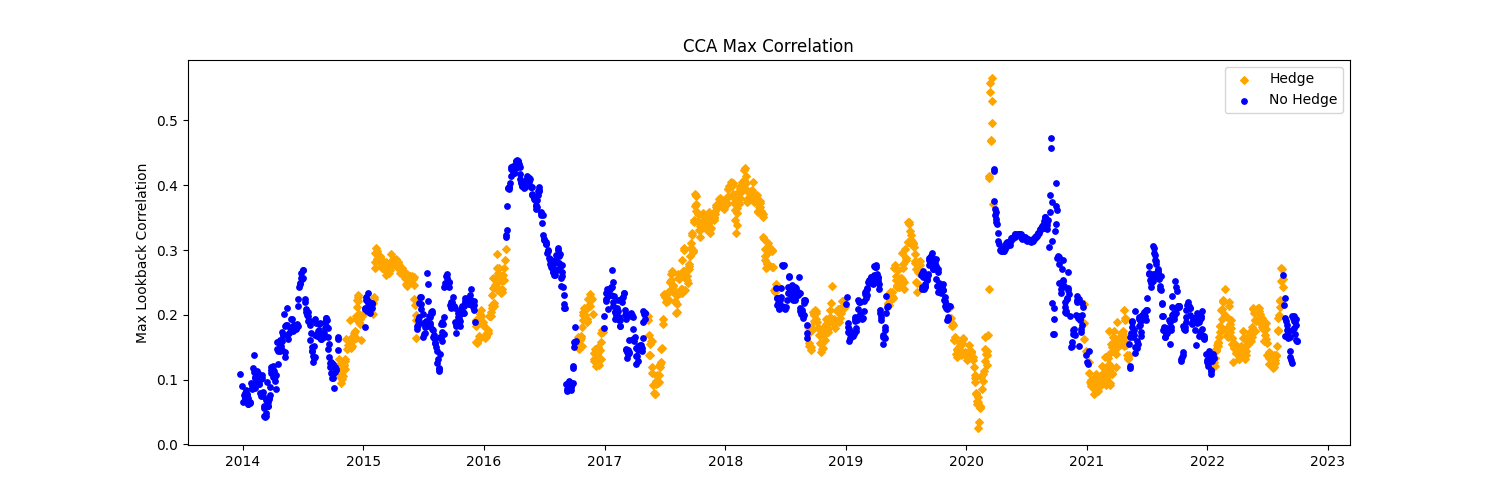}
  \caption{Correlation and Hedge Timing Produced by the CCA for PIMIX Hedged with LQD only and Considering All Signals.}
  \label{LQD CCA Correlation}
\end{figure}

\subsection{Credit Default Swap Index: IG CDX}
Figure \ref{LQD_All_no_costs} above displayed that LQD is a very effective hedge, presumably due to IG bond convexity in drawdowns. An alternative hedge instrument considered are IG credit default swaps (IG CDXs). IG CDXs have deep liquidity, low trading and funding costs, and could be highly effective in providing swift and exhaustive downside coverage for a credit portfolio without fears of strictly binding volume constraints and market impact. However, credit default swaps do not benefit from the same downside convexity that credit ETFs possess.

\begin{figure}
    \caption{$\bold{CCA \ Hedged \ Portfolio}$ $\bold{PIMIX}$, $\bold{IG \ CDX \ Hedge}$, $\bold{All \ Signals}$, $\bold{No \ Costs}$.}
    \centering
    \begin{subfigure}{\linewidth}
        \centering
        \includegraphics[width=\linewidth]{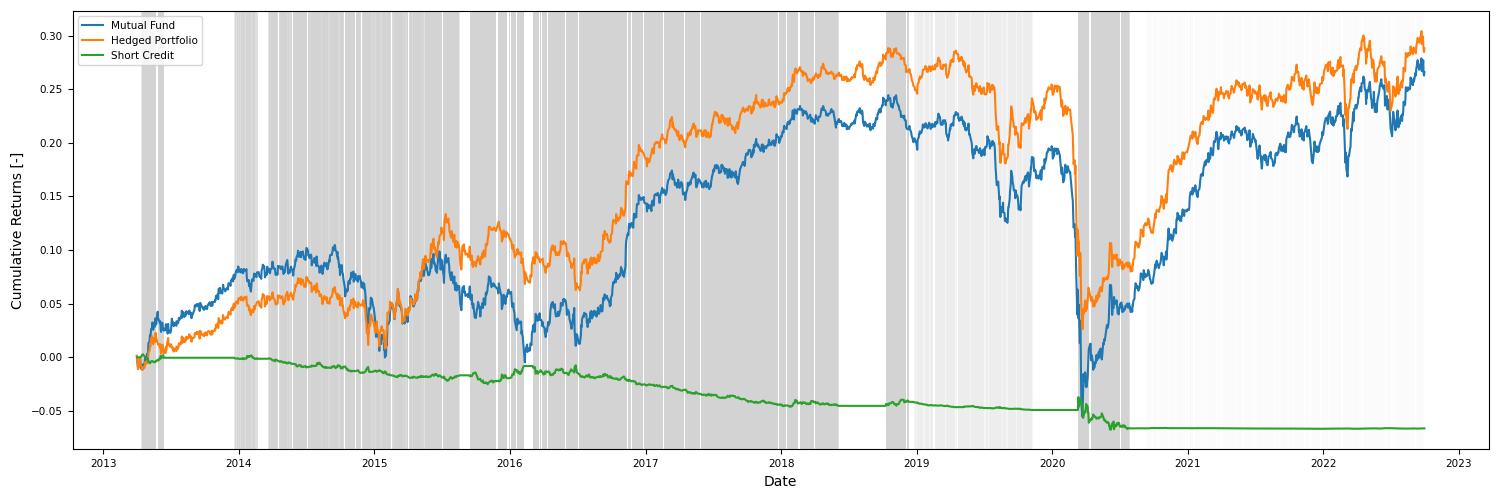}
        \caption{Cumulative Returns}
        \label{IGCDX_combined_returns}
    \end{subfigure}

    \vspace{0.5cm} 
    
    \begin{subfigure}{\linewidth}
        \centering
        \includegraphics[width=\linewidth]{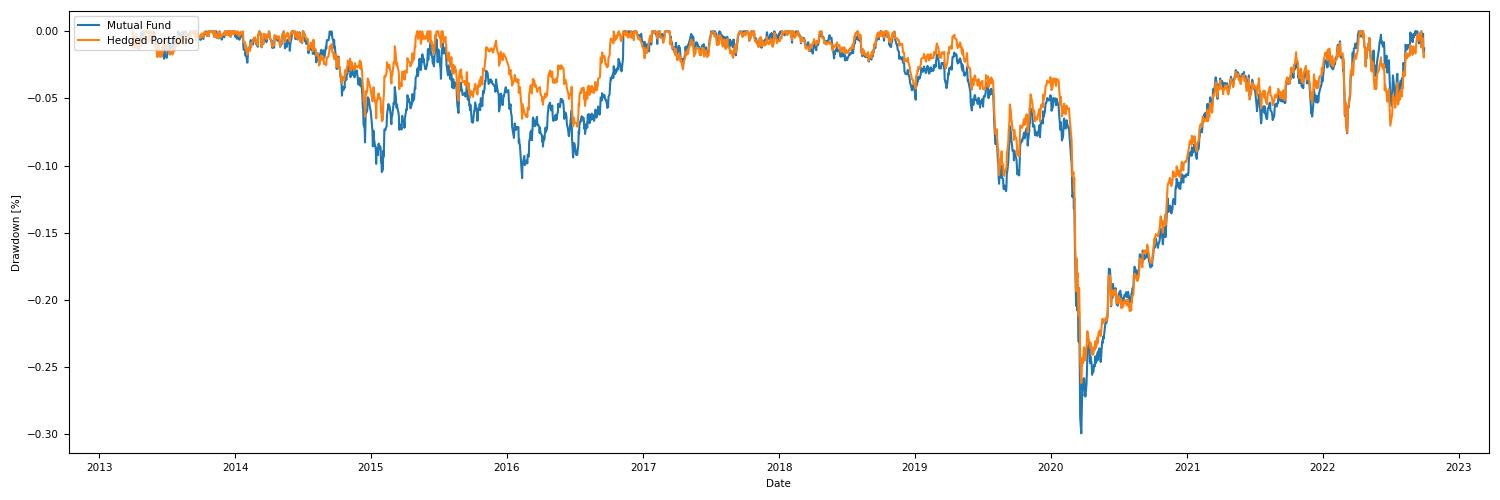}
        \caption{Drawdowns}
        \label{IGCDX_combined_drawdowns}
    \end{subfigure}

    \vspace{0.5cm} 
    
    \begin{subfigure}{\linewidth}
        \centering
        \includegraphics[width=\linewidth]{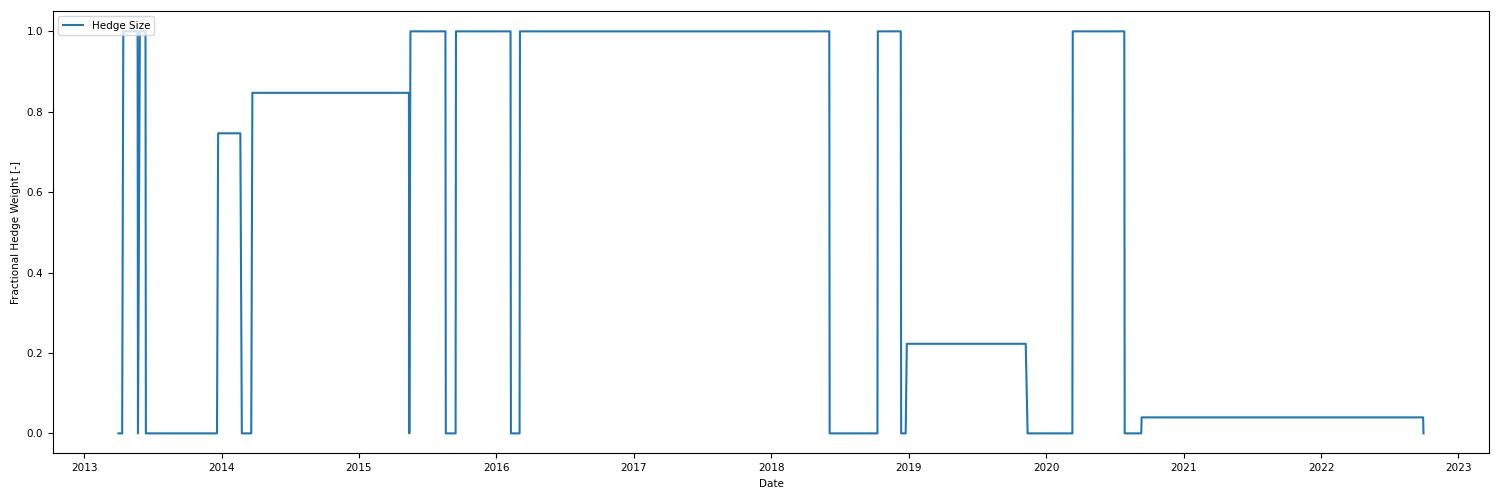}
        \caption{Hedge Weights and Activations}
        \label{IGCDX_combined_activations}
    \end{subfigure}

    \label{IGCDX_All_no_costs}
\end{figure}

Figure \ref{IGCDX_All_no_costs} displays the IG CDX's CCA informed hedge performance, considering all three signals (Credit, Liquidity and Momentum). Note that even though the hedge is active in several market downturns, the hedge does not perform nearly as well as with LQD. In addition, the timings for entries and exits are not as precise.

\clearpage

\begin{figure}[H]
  \centering
  \includegraphics[width=0.675\linewidth]{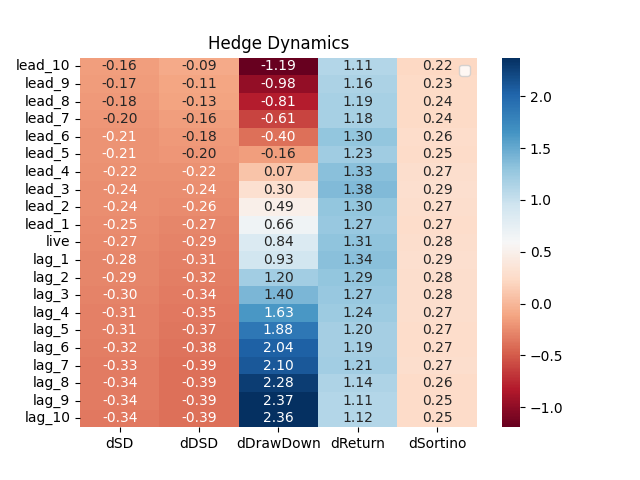}
  \caption{CCA Hedged PIMIX, IG CDX Hedge, All Signals, No Costs.}
  \label{IG_CDX Risk Metrics}
\end{figure}

As evident in Figure \ref{IG_CDX Risk Metrics} though shorting IG CDX does generate net returns outperformance, the hedge does not cancel out the 2020 drawdown despite the hedge being active. The assertion that credit default swaps lack downside convexity and did not provide much protection during 2020 holds in the data. IG CDXs alone do not suffice for avoiding large drawdowns, and make minimal contribution to Sortino gains. Figure \ref{IG_CDX Risk Metrics} further reflects that CDX timing is not as precise, and that even leading the signals does not help avoid drawdowns or increase the Sortino Ratio when hedging the portfolio with only IG CDX. Consequently, the focus of the following sections are on introducing further credit ETFs as hedges, and only supplementing hedge liquidity with CDXs for faster implementation when it comes to very large portfolio sizes.

\clearpage

\subsection{Hedged Performance Considering Transaction Costs, Funding Costs and Portfolio Size}

Funding costs for borrowing instruments such as LQD can vary from as low as 20bps to as much as 800bps on rare days with large redemptions. Empirical funding costs show that the funding cost is relatively insensitive to market downturns. As such, the following section analyzes the performance of the CCA LQD hedged portfolio as defined in Eq. (\ref{Costs1}) for various short funding costs and when considering all three signals.

Table \ref{LQD Considering Costs} showcases results that confirm the expected inverse relationship between funding cost and hedged portfolio outperformance, with higher hedge funding costs driving the hedged portfolio's gain in Sortino lower.
Interestingly, the performance of the hedged portfolio is still quite good (compared to the benchmark), even for quite large funding costs ($f_{bps} \geq 100$). In addition, note that the ``optimal" tuned parameters (based on the three-step optimization displayed in the previous section) don't change until $f_{bps} = 200$, whereby the algorithm determines it's more beneficial to turn the hedge off at a lower (in magnitude) threshold ($\gamma_{CCA, lower} = -1.0$ rather than $\gamma_{CCA, lower} = -1.5$).

\begin{table}[H]
    \centering
    \caption{Tuned Parameters and Increase in Sortino Ratio for the Hedged Portfolio when Considering Transaction Costs and for Various Funding Costs.}
    \begin{tabular}{>{\centering\arraybackslash}m{1.0in}>{\centering\arraybackslash}m{0.75in}>{\centering\arraybackslash}m{0.75in}>{\centering\arraybackslash}m{0.75in}>{\centering\arraybackslash}m{1.25in}}
    \toprule
    Funding Cost, $f_{bps}$ [BP] & Lookback [Days] & $\gamma_{CCA,lower}$ [std] & $\gamma_{CCA,upper}$ [std] & $\Delta \text{Sortino Ratio}_{252}$ $\text{ [std}^{-1}]$\\
    \midrule
    0 & 40 & -1.50 & 2.50 & 0.787 \\
    20 & 40 & -1.50 & 2.50 & 0.755 \\
    50 & 40 & -1.50 & 2.50 & 0.732 \\
    100 & 40 & -1.50 & 2.50 & 0.694 \\
    200 & 40 & -1.00 & 2.50 & 0.618 \\
    \bottomrule
    \end{tabular}
    \label{LQD Considering Costs}
\end{table}

Next, building on the transaction cost and funding costs included right above, realistic hedge implementation requires consideration of the bond portfolio's size relative to LQD's traded volume to inform gradual position taking with minimal market impact. To this end, the following considers the performance of the above CCA, LQD hedged portfolio with funding cost set to ($f_{bps} = 50$) and three different fund sizes 100 Million, 1 Billion and 10 Billion USD (proxies for small, medium and large sized bond funds). This approach invokes the volume-based transactions limitation described in Eq. (\ref{Costs3}) (namely that the manager can only short up to 10\% of the 252 day simple-moving-average of the instrument trading volume). For illustration, consider the scenario where the fund notional value is 10 Billion USD, the CCA recommends the full fund be hedged by LQD with a spot price of \$100.0 USD and having $V_{t,252} = 1,000,000$. It would take the fund manager 1,000 trading days just to put the hedge on in this one instance.


Table \ref{LQD Considering Cost and Volume} displays the hedged portfolio performance for the three different fund sizes (\$0.1bln, \$1bln and \$10bln USD). For small funds to medium sized funds (\$0.1bln, \$1bln) that require moderate volume taking, hedged performance and model parameters remain consistent. As the fund size increases to a large fund (\$10bln), demanded liquidity is substantial, and the model prefers longer lookbacks and higher turn on thresholds, meaning the hedge slows down. This volume-informed framework produces a consistent pairing between growing fund size, slower hedging, and lower hedged performance.



\begin{table}[H]
    \centering
    \caption{Tuned Parameters and Increase in Sortino Ratio for the Hedged Portfolio when Considering Transaction and Funding Costs and for Various Fund Sizes.}
    \begin{tabular}{>{\centering\arraybackslash}m{1.0in}>{\centering\arraybackslash}m{0.75in}>{\centering\arraybackslash}m{0.75in}>{\centering\arraybackslash}m{0.75in}>{\centering\arraybackslash}m{1.25in}}
    \toprule
    Fund Size [Millions \$] & Lookback [Days] & $\gamma_{CCA,lower}$ [std] & $\gamma_{CCA,upper}$ [std] & $\Delta \text{Sortino Ratio}_{252}$ $\text{ [std}^{-1}]$\\
    \midrule
    100 & 40 & -1.5 & 2.5 & 0.744 \\
    1,000 & 40 & -1.5 & 2.0 & 0.437 \\
    10,000 & 125 & -1.5 & 3.0 & 0.223 \\
    \bottomrule
    \end{tabular}
    \label{LQD Considering Cost and Volume}
\end{table}

\begin{figure}
    \caption{$\bold{CCA \ Hedged \ Portfolio}$ $\bold{PIMIX}$, $\bold{LQD \ Hedge}$, $\bold{All \ Signals}$, $\bold{All \ Costs}$.}
    \centering
    \begin{subfigure}{\linewidth}
        \centering
        \includegraphics[width=\linewidth]{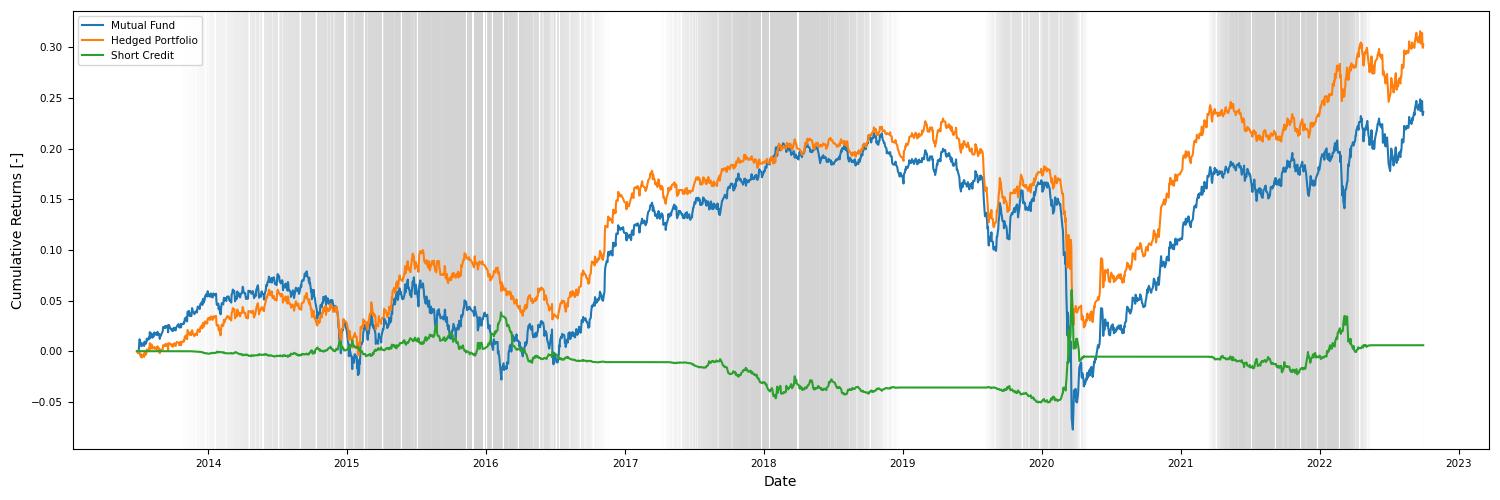}
        \caption{Cumulative Returns}
        \label{LQD_all_costs_returns}
    \end{subfigure}

    \vspace{0.5cm} 
    
    \begin{subfigure}{\linewidth}
        \centering
        \includegraphics[width=\linewidth]{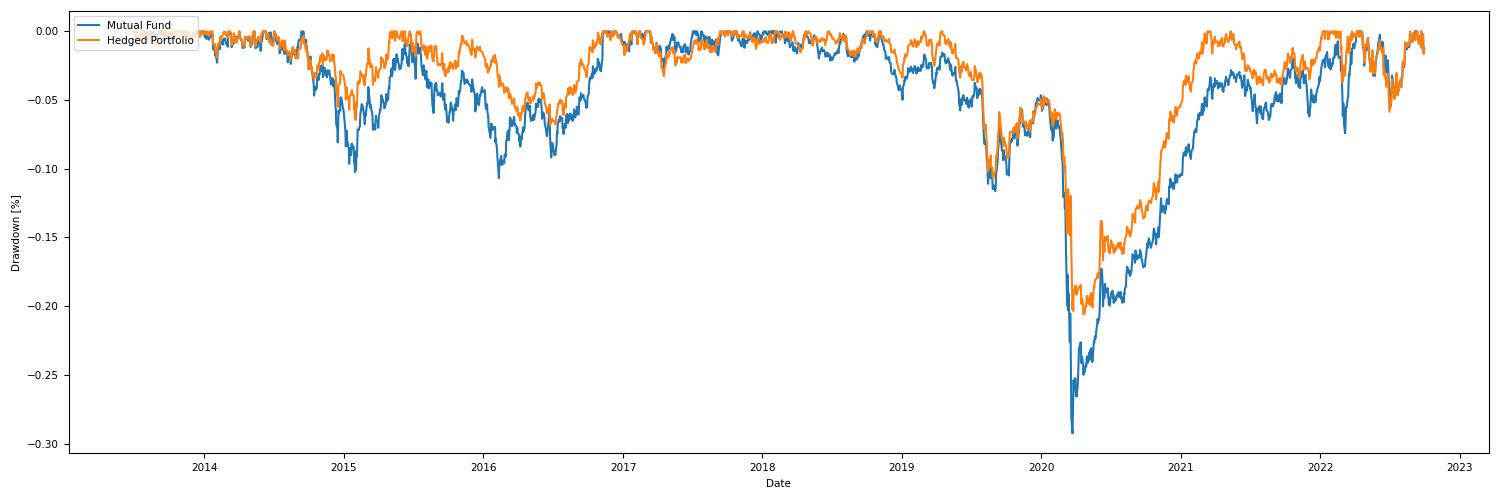}
        \caption{Drawdowns}
        \label{LQD_all_costs_drawdowns}
    \end{subfigure}

    \vspace{0.5cm} 
    
    \begin{subfigure}{\linewidth}
        \centering
        \includegraphics[width=\linewidth]{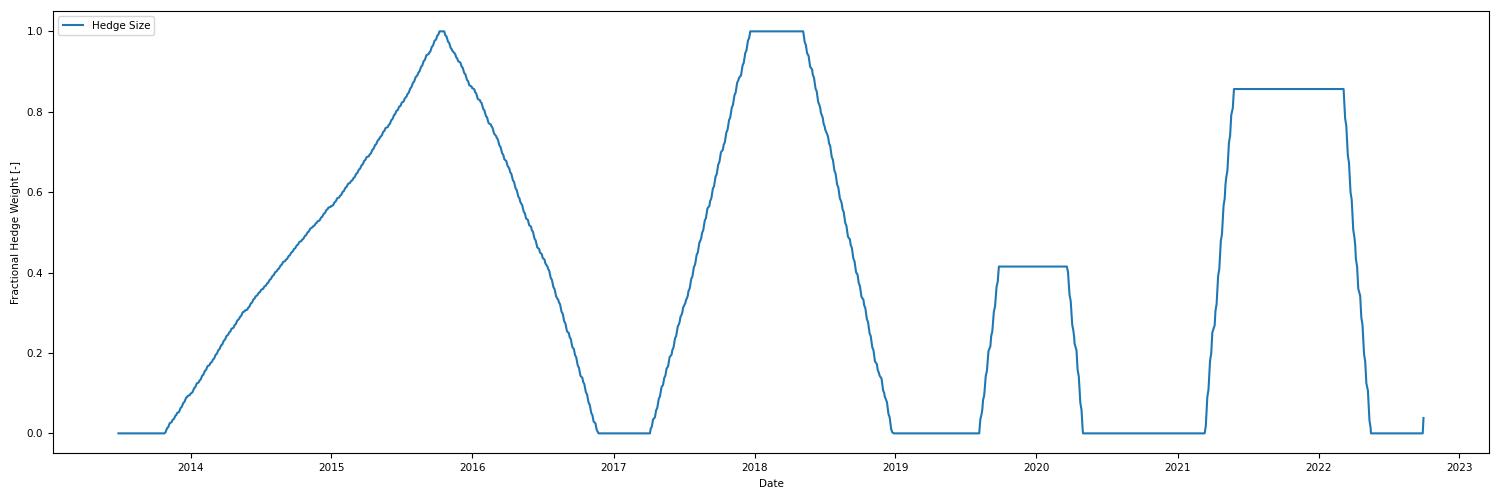}
        \caption{Hedge Weights and Activations}
        \label{LQD_all_costs_activations}
    \end{subfigure}
    \label{LQD_All_costs}
\end{figure}

Figure \ref{LQD_All_costs} displays the cumulative returns, drawdowns and hedge weights for the the large (\$10bln) fund, considering transaction and funding costs ($f_{bps} = 50$) and with all three signals. As noted, the cumulative returns of the hedged portfolio are not as good as those displayed in Figure \ref{LQD_All_no_costs}. Most interestingly, Figure \ref{LQD_all_costs_activations} shows that the hedge only activates four times during the nearly 11 year window and takes \textbf{more than two years} to activate starting at the end of 2013. Turning the hedge on over a two year window involves shorting LQD every trading day during this period of time and is most likely unrealistic. 

This result shows that for funds in excess of 1 Billion USD, shorting LQD \textbf{only} as a tactical hedge is likely not viable. Further analysis reveals there is a bottleneck in hedge speed and performance for funds around \$5bln or greater in assets. With this acknowledged, the following section considers adding additional hedge instruments,
to distribute demanded liquidity to multiple hedge ETFs (LQD, HYG, JNK) and also to CDXs as highly liquid albeit less effective (due to lower downside convexity) alternative hedges. 


\clearpage

\section{Hedging PIMIX with Multiple Instruments}\label{sec:Multiple Hedges}

Having established that hedge liquidity is a binding constraint for funds above a threshold of \$5bln, the following section includes more ETFs as hedge instruments in order to distribute liquidity. CCA full cost (i.e. considering trading and funding costs with $f_{bps} = 50$) backtests determine the hedge effectiveness for a fund size of \$10bln, using all signals, and introducing HYG as an additional hedge. Note that HYG likely has larger negative carry during calm market times but adds even more downside convexity to the hedged portfolio, potentially increasing performance during drawdowns.

\begin{figure}
    \caption{$\bold{CCA \ Hedged}$ $\bold{PIMIX}$, $\bold{LQD \ \& \ HYG \ Hedges}$, $\bold{All \ Signals}$, $\bold{All \ Costs}$.}
    \centering
    \begin{subfigure}{\linewidth}
        \centering
        \includegraphics[width=\linewidth]{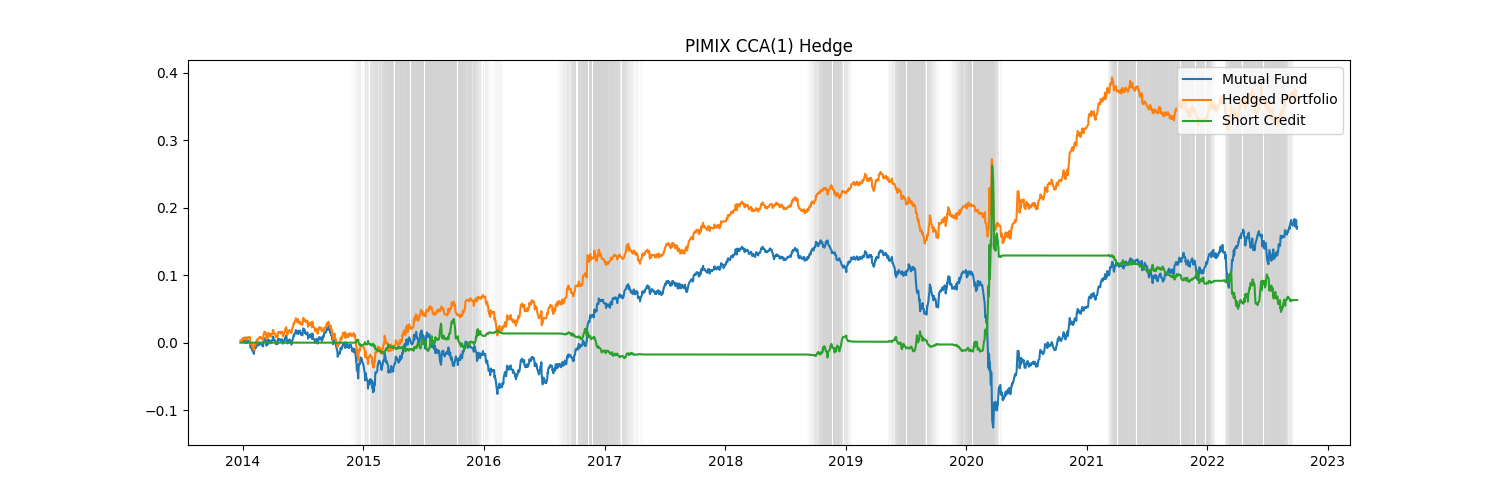}
        \caption{Cumulative Returns}
        \label{LQD_all_costs_returns}
    \end{subfigure}

    \vspace{0.5cm} 
    
    \begin{subfigure}{\linewidth}
        \centering
        \includegraphics[width=\linewidth]{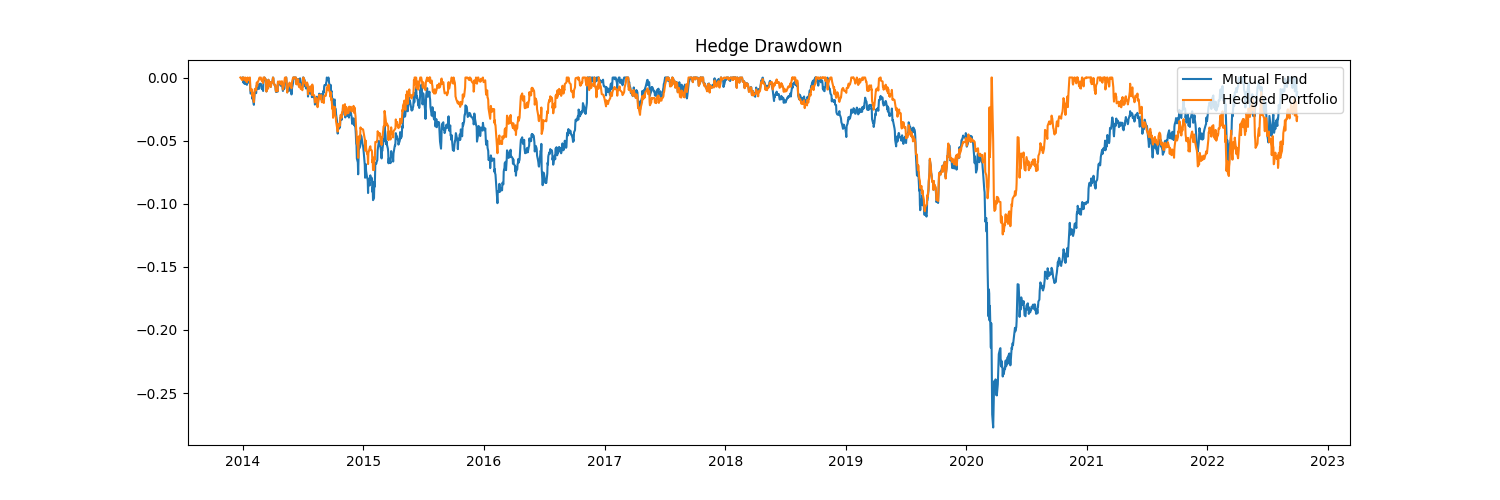}
        \caption{Drawdowns}
        \label{LQD_all_costs_drawdowns}
    \end{subfigure}

    \vspace{0.5cm} 
    
    \begin{subfigure}{\linewidth}
        \centering
        \includegraphics[width=\linewidth]{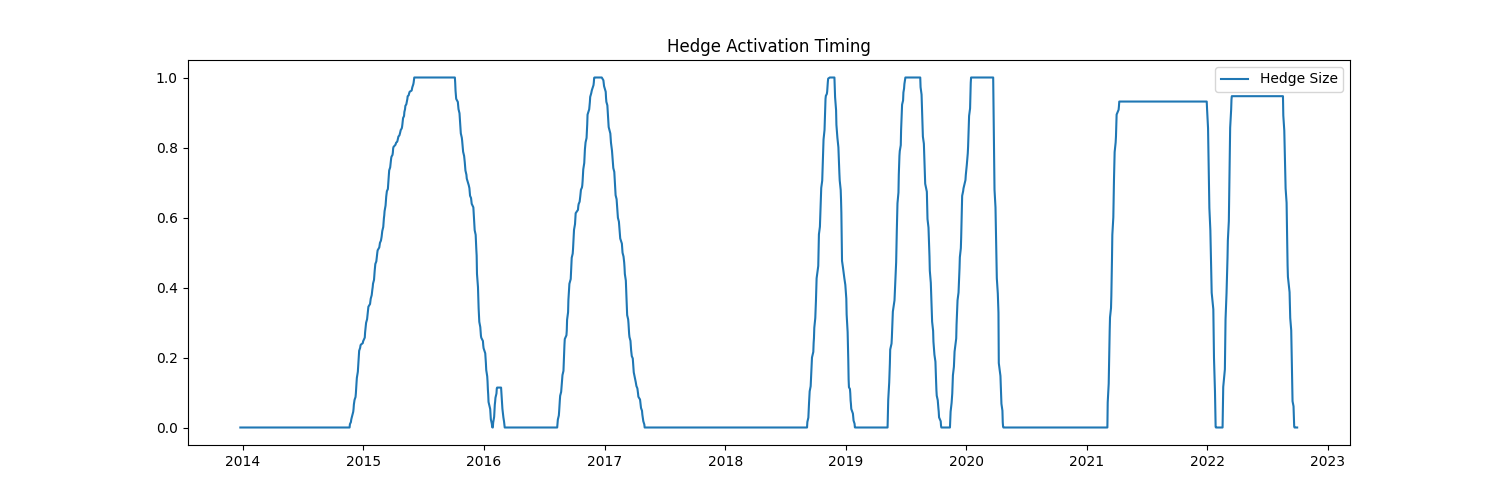}
        \caption{Hedge Weights and Activations}
        \label{LQD_all_costs_activations}
    \end{subfigure}
    \label{LQD_Multi_hedge}
\end{figure}

Figure \ref{Hedge Volumes} shows that HYG trades (on average) three-times more shares than LQD and should provide more shares to short, hopefully increasing the hedge activation and deactivation. Figure \ref{LQD_Multi_hedge} shows the performance of the hedge with LQD and HYG as hedge instruments. Note that indeed, the hedge appears significantly faster now, activating and deactivating within a matter of months, and successfully capturing multiple drawdowns. The inclusion of HYG as a hedge instrument brings the hedged portfolio performance in line with that of the hedged medium sized funds (i.e., \$5bln), enhancing Sortino (compared to the baseline portfolio) by around 0.4. Note that another backtest (not shown) which included IG CDX as another hedge instrument (in addition to LQD \& HYG) only marginally improved the hedged portfolio performance due its lack of downside convexity (mentioned in Section \ref{sec:Single Hedge}).

Considering the hedge performance (cumulative returns and drawdows) from Figure \ref{LQD_all_costs_returns}-\ref{LQD_all_costs_drawdowns}, it's clear that the strategy works well overall, significantly outperforming the benchmark; however, there are certain periods of time when the hedge is on, but the hedged portfolio draws down just as much as the benchmark. This behavior occurs most notably during late 2019, where both funds draw down approximately 11\%. The figures show that the hedge picks up some benefit during the drawdown, but the nature of the drawdown is more prolonged (rather than very sharp), and even with a slight rally during its midst. This rally hurts the hedged portfolio performance as does the extended nature, which induces higher funding costs. No hedge can be 100\% effective, but the results show that the current hedging strategy is indeed largely beneficial for funds such as PIMIX.

\clearpage

\begin{figure}[H]
  \center
  \includegraphics[height=0.4\textheight,
  width=1\textwidth]{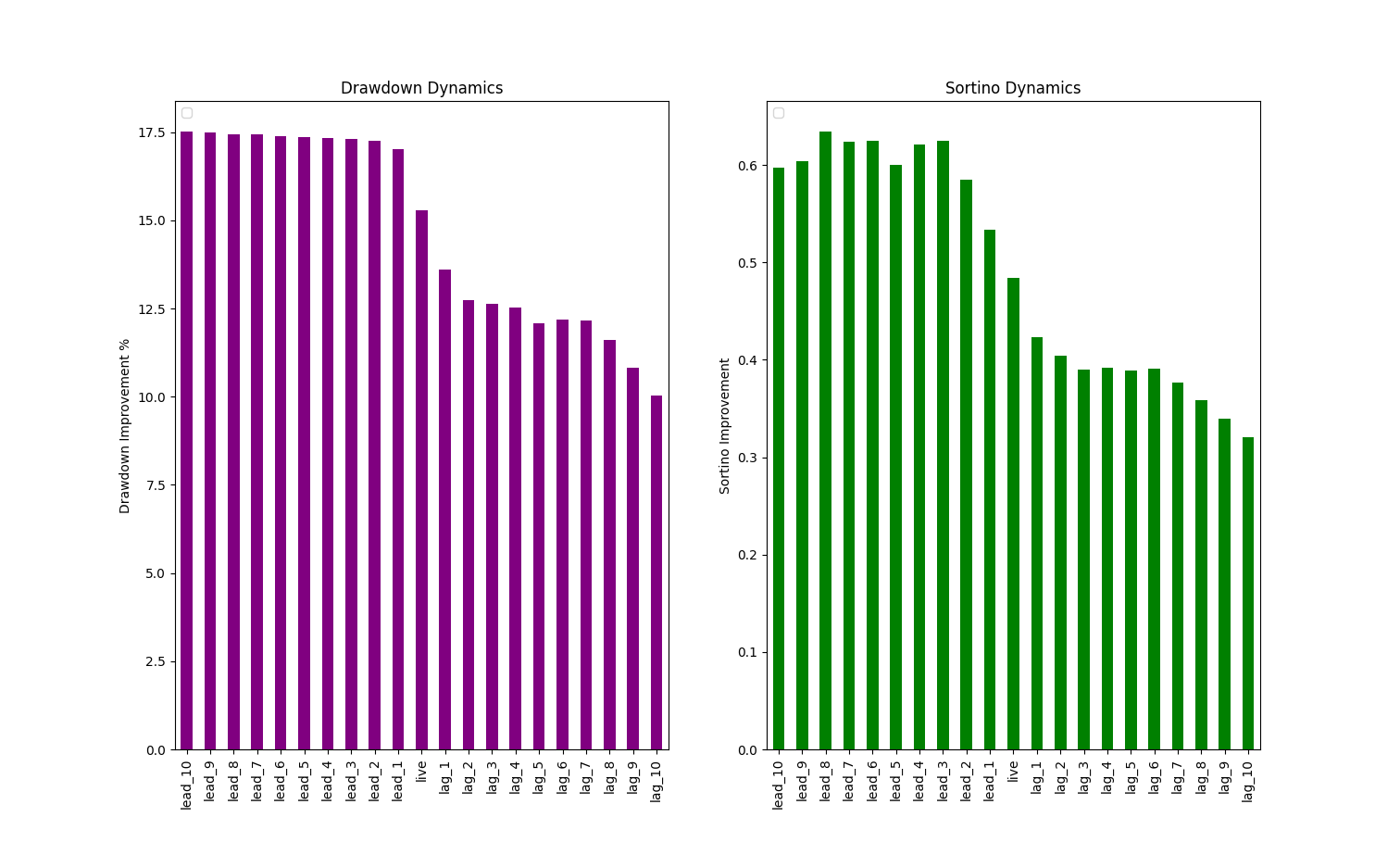}
  \caption{Pimix, LQD \& HYG Hedge, Drawdown and Sortino Lagged Dynamics}
  \label{LQD_dynamics}
\end{figure}

Figure \ref{LQD_dynamics} shows that the hedge's performance does not depend on instantaneous action. Similar performance is attainable even if the CCA hedge is operated with up to two business weeks of delay. As expected, leading the signals increases the effectiveness of the strategy, and acting with the delay reduces performance - demonstrating there is valuable information content in the signals.

\clearpage

\section{Hedging Other Funds: DODIX}\label{sec:DODIX}

This section shifts focus to evaluating how generalisable the results learned from the canonical correlation backtests from the previous section are to other funds. The framework so far indicates the existence of a relatively cheap hedge that assists in avoiding drawdowns and increases annualised returns on PIMIX. PIMIX is a fund whose mandate reflects relatively high active risk with higher spread volatility and a large illiquidity premium. In contrast, DODIX appears more conservative, with lower and more stable spread returns over the last decade, as evident in Figure \ref{pimix_vs_dodix} below.

\begin{figure}[H]
  \center
  \includegraphics[height=0.45\textheight,
  width=0.9\textwidth]{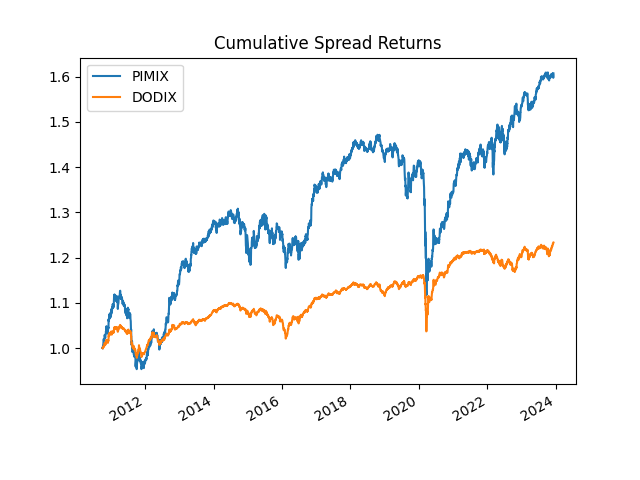}
  \vspace{-10mm}
  \caption{PIMIX \& DODIX Cumulative Spread Returns}
  \label{pimix_vs_dodix}
\end{figure}

The case of DODIX highlights the importance of volatility scaling via the ratio between the fund's and the hedges' historical duration neutral return volatility becomes valuable (see end of Chapter 4). Without such scaling, the dynamic hedge for DODIX would actually be a  net tactical short on LQD. Since DODIX targets lower spread return, a portfolio manager seeking to hedge DODIX would require hedge instruments with lower carry, as the fund has less carry to ``burn". As such, the hedges tested on DODIX are LQD and IG CDX, as HYG and HY CDX as hedges are too expensive in this scenario.

\clearpage

\begin{figure}[H]
  \center
  \includegraphics[height=0.3\textheight,
  width=1\textwidth]{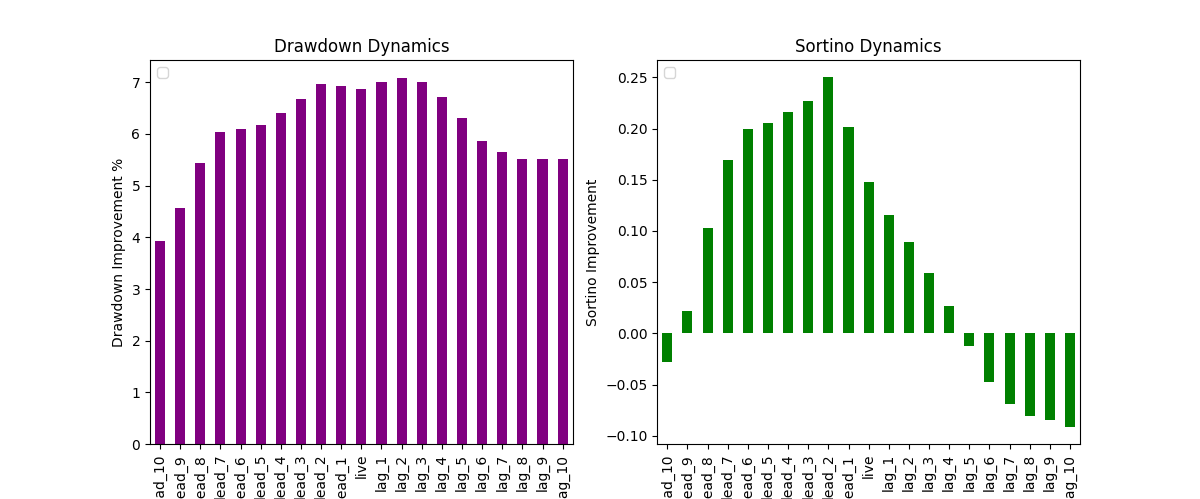}
  \vspace{-10mm}
  \caption{DODIX Drawdown \& Sortino Dynamics}
  \label{dodix_dynamics_1}
\end{figure}

\begin{figure}
    \caption{$\bold{CCA \ Hedged} \ \bold{DODIX}$, $\bold{LQD \ \& \ IG \ CDX \  Hedges}$, $\bold{All \ Signals}$, $\bold{All \ Costs}$.}
    \centering
    \begin{subfigure}{\linewidth}
        \centering
        \includegraphics[width=\linewidth]{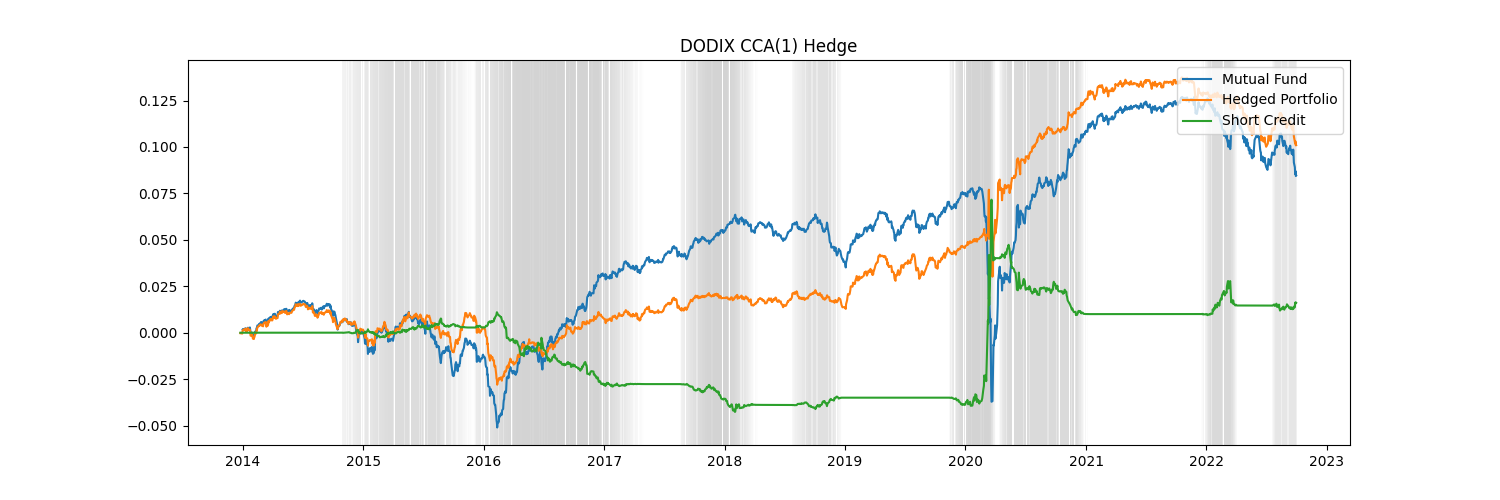}
        \caption{Cumulative Returns}
        \label{DODIX_all_costs_returns}
    \end{subfigure}

    \vspace{0.5cm} 
    
    \begin{subfigure}{\linewidth}
        \centering
        \includegraphics[width=\linewidth]{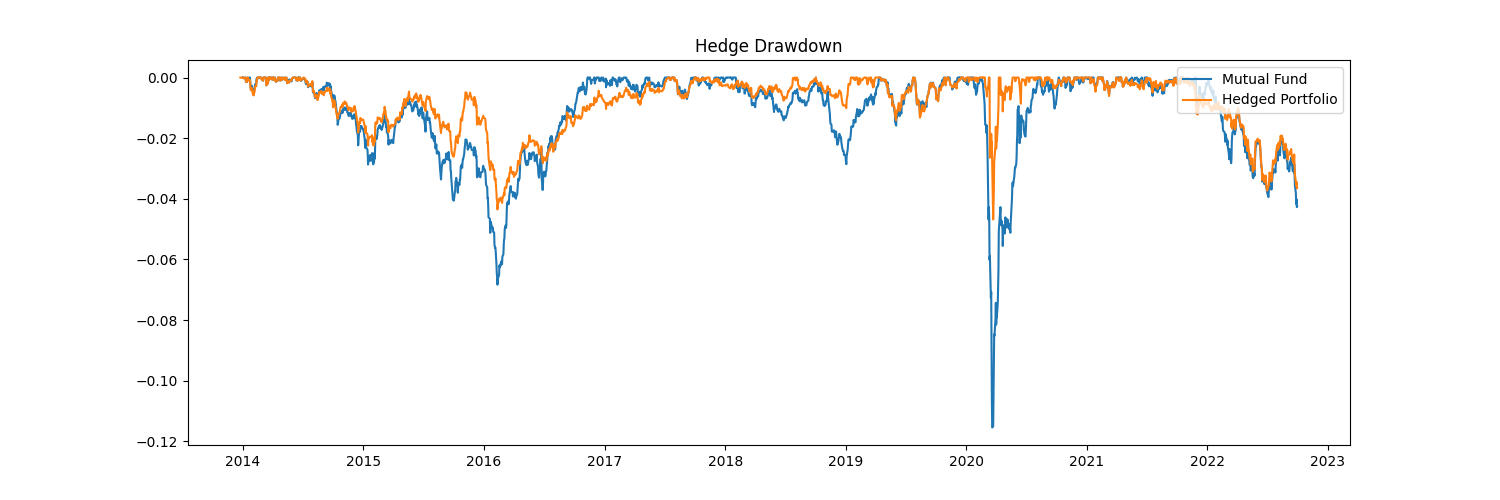}
        \caption{Drawdowns}
        \label{DODIX_all_costs_drawdowns}
    \end{subfigure}

    \vspace{0.5cm} 
    
    \begin{subfigure}{\linewidth}
        \centering
        \includegraphics[width=\linewidth]{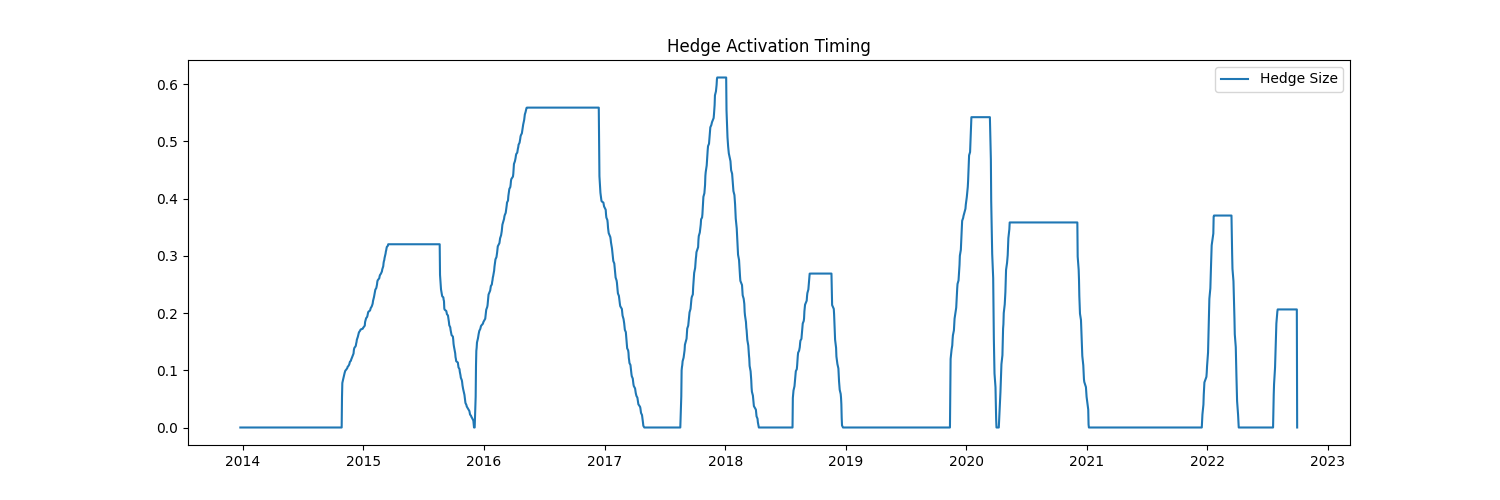}
        \caption{Hedge Weights and Activations}
        \label{DODIX_all_costs_activations}
    \end{subfigure}
    \label{DODIX_all}
\end{figure}

Figure \ref{DODIX_all} shows the results for a CCA informed hedge on a \$10bln conservative fund (DODIX returns used as a proxy) with all three signals (Credit, Liquidity and Momentum), utilizing LQD and IG CDX as hedges and incorporating realistic position taking, bid-ask spreads, and funding costs of $f_{bps}=20$. Note that the ideal parameters for the fund are: Lookback = 60 days, $\gamma_{CCA,lower}=-1.50 \text{ and } \gamma_{CCA,upper}=2.50$, very similar to those for PIMIX. Evidently volatility scaling now binds, with the maximum hedge weight below 0.6. Even so, the negative carry of the hedges is quite expensive between 2016-2019, before eventually paying off in 2020. The majority of DODIX's $\geq 10\%$ drawdown is neutralised, if the hedge is initiated within a two week action window. DODIX's gain in Sortino is marginal, lower, and decays much faster than PIMIX's gain in Sortino, as displayed in in \ref{dodix_dynamics_1}.

\clearpage

\begin{figure}[H]
  \center
  \includegraphics[height=0.35\textheight,
  width=0.7\textwidth]{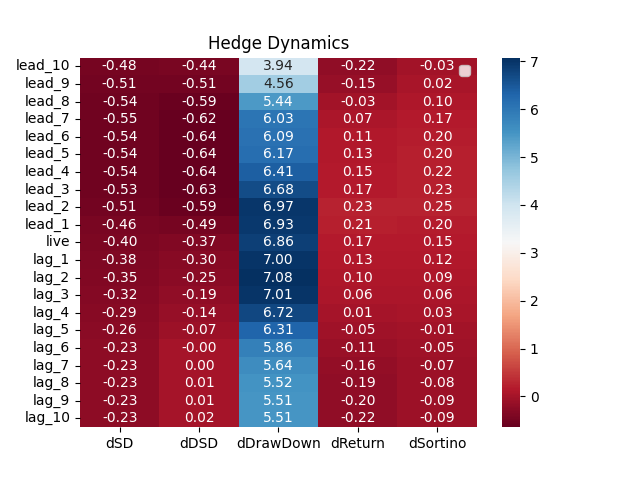}
  \vspace{-10mm}
  \caption{DODIX Hedge Dynamics}
  \label{dodix_dynamics_2}
\end{figure}

DODIX's CCA hedge effectively lowers standard deviation, reduces downside standard deviation, and scales back the 2016 and 2020 drawdowns - though it flattens returns and therefore has limited impact on Sortino, as seen in \ref{dodix_dynamics_2}. Overall, CCA's results are fairly strong. It does appear that the liquidity premium that comes from shorting credit ETFs offers is most suitable and a great match to higher carry, more illiquid, high active risk funds like PIMIX. Still, CCA offers great downside protection on lower carry and lower active risk funds as in DODIX, when the CCA hedge is effectively paired with volatility scaling.

\clearpage

\clearpage
\chapter{Conclusions} \label{Ch. 6 - Conclusions}




The research above constructed signals on the Investment Grade bond market to inform a dynamic hedge that deploys liquid bond ETFs as hedges to effectively and quickly protect high carry bond funds. It succeeded in lowering absolute and relative risk, increasing annualised returns, and improving Sortino for PIMIX and avoiding drawdowns for DODIX, in a realistic framework that incorporates trading costs, funding costs, and volume sized hedge positions.

Credit Risk, Liquidity, and Momentum signals derived from options, duration times spread, and cumulative duration-neutral returns respectively, each seemed to capture some orthogonal information about the IG bond market. Hedge performance considering individual signals, followed by their combination, proves this point - with an optimal improvement in Sortino of $\geq 0.7$ using the joint signals. When searching the hedge model's parameter space, results remain strong and consistent over a wide array of tested parameters.

The Canonical Correlation (CCA) optimisation dynamic hedging method learns by maximising the correlation of signals to hedged returns. Table \ref{LQD Tuned Parameters} shows CCA favours medium horizon credit and liquidity risk combined with short-term momentum, presumably loading on credit and liquidity risk shocks to enter the hedge and short-term reversal to exit. It is clear in figure \ref{LQD CCA Correlation} that CCA learns to activate the hedge when correlations from signals to hedged returns begin to spike, and exit at the peak or when correlations revert back to normal levels. This benefits bond portfolio managers by offering great empirical downside protection, limiting drawdown risk while keeping hedging costs low.

Hedging is cost effective as the research has focused on establishing short positions in IG (LQD) and HY (HYG) bond ETFs rather than shorting individual IG corporate bonds. IG bond ETFs are liquid and have low bid-ask spreads, and establishing shorts in the IG bond ETF space via LQD \& HYG provides great downside convexity which benefits the efficacy of the hedge. While IG and HY CDXs have far larger traded volumes than LQD \& HYG, they do not have the same downside convexity and prove to be not as effective as ETFs.

The backtesting framework applied on duration neutral returns of PIMIX and DODIX from July 2013 - October 2022 yielded results that showed high carry funds like PIMIX can effectively be hedged by shorting a PCA weighted combination of LQD, HYG and IG CDX, even when considering realistic volume taking, transaction costs, and funding costs for the hedge. When the portfolio size (i.e. notional value) exceeds \$5bln USD, it is necessary to utilize multiple hedge instruments to hedge the portfolio within a reasonable time. An extension of the hedge to DODIX, a more conservative fund with lower carry, outlines the hedge's robustness and downside protection using LQD and IG CDX, without sacrificing much upside. The backtested hedges also withstand lags to implementation and are not contingent on instantaneous action, making them favourable.


The research herein represents a significant finding which should be applicable (and implementable) for managers of small (less than \$1bln) to moderately large (\$10bln) bond funds. With the global growth of the ETFs, it stands to reason that transacted volumes on these instruments will continue to increase (see the trend in Figure \ref{Hedge Volumes}), increasing the effectiveness and viability of this dynamic hedge. 

\clearpage
\bibliographystyle{plain}
\bibliography{main}
\clearpage

\appendix
\chapter*{Appendix}\label{Remove Duration}
\addcontentsline{toc}{chapter}{Appendix:  Removing Duration from Portfolio \& Hedge Instrument Returns}
\counterwithin{equation}{section}
\setcounter{equation}{0}
\renewcommand\theequation{A\arabic{equation}}

As mentioned in the body of the manuscript, IG and HY bonds prices (and therefore returns) depend on their spreads to US Treasury bonds with similar durations. The first step in removing duration effects from the corporate bond ETF and portfolio returns is to identify the Treasuries with durations that closely match that of the instrument. Specifically, given a target asset with duration $D_{target}$, and a treasury universe with durations $D_1, ..., D_n$, select two treasuries whose durations ($D_{i^\ast}$ and $D_{j^\ast}$) bound $D_{target}$, i.e.

\begin{equation}\label{A1}
    \begin{aligned}
        D_{i^\ast} &=\min_{{0\le i\le n}} {{(D}_i-D_{target}|}D_i>D_{target}) \\
        D_{j^\ast} &=\min_{{0\le j\le n}} {{(D}_{target}-D_j|}D_j<D_{target})
    \end{aligned}
\end{equation}

A simple linear combination of these two Treasuries then matches the target duration $D_{target}$, with the constraint that their weights sum to one.

\begin{equation}\label{A2}
    \begin{aligned}
        w_{i^\ast} &= \frac{D_{target} - D_{j^\ast}}{D_{i^\ast} - D_{j^\ast}} \\
        w_{j^\ast} &= 1 - w_{i^\ast}         
    \end{aligned}
\end{equation}

\noindent The duration-driven (simple) return (i.e. the weighted average of $R_{i^\ast}$ and $R_{j^\ast}$) is then

\begin{equation}\label{A3}
    R_{duration}=w_{i^\ast}R_{i^\ast}+w_{j^\ast}R_{j^\ast}    
\end{equation}

\noindent Finally, the duration-neutral return is

\begin{equation}
  \Tilde{R}_{target}=R_{target}-R_{duration}  
\end{equation}

\end{document}